\pdfoutput=1
\documentclass[10pt, conference, compsocconf]{IEEEtran}
\usepackage{cite}
\usepackage{adjustbox}
\usepackage{graphicx}
\usepackage{booktabs}
\usepackage{multirow}
\usepackage{multicol}
\usepackage{amsmath}
\usepackage{amssymb}
\usepackage{subcaption}
\usepackage{ifthen}
\usepackage{ifpdf}
\usepackage{color}
\usepackage{url}
\usepackage{array}
\usepackage{tabularx}
\let\labelindent\relax
\usepackage{enumitem}
\usepackage{comment}
\usepackage{algorithm}
\usepackage{algorithmicx}
\usepackage{algpseudocode}
\usepackage{caption}
\usepackage{lipsum}
\usepackage{siunitx}
\usepackage{fancyvrb}
\usepackage{listings}

\makeatletter
\newcommand{\verbatimfont}[1]{\renewcommand{\verbatim@font}{\ttfamily#1}}
\makeatother

\usepackage{tgcursor}

\def\IEEEbibitemsep{0pt plus .5pt}

\newcommand{\multiletterSymbol}[1]{\ensuremath{\mathit{#1}}}
\renewcommand{\TH}{\multiletterSymbol{TH}}
\newcommand{\FV}{\multiletterSymbol{FV}}
\newcommand{\BV}{\multiletterSymbol{BV}}

\definecolor{rd} {rgb} {1.0,0.0,0.0}
\definecolor{lg} {rgb} {0.7,0.3,0.9}
\definecolor{db} {rgb} {0.0,0.0,0.7}
\definecolor{dg} {rgb} {0.0,0.7,0.0}
\definecolor{dr} {rgb} {0.7,0.0,0.0}
\definecolor{gr} {rgb} {0.8,0.4,0.1}

\newcommand{\John } [1] {{{\color{db}[John] #1}}}
\newcommand{\Roger} [1] {{{\color{dg}[Roger]  #1}}}
\newcommand{\YC   } [1] {{{\color{lg}[YC]#1}}}

\newcommand{\final} {1}

\ifthenelse{\equal{\final}{1}} {\renewcommand{\John }[1]{}}{}
\ifthenelse{\equal{\final}{1}} {\renewcommand{\Roger }[1]{}}{}
\ifthenelse{\equal{\final}{1}} {\renewcommand{\YC }[1]{}}{}

\providecommand{\shortcite}[1]{\cite{#1}}
\ifdefined\textln\relax\else\newcommand{\textln}[1]{{\fontfamily{pplx}\selectfont #1}}\fi
\makeatletter
\newenvironment{breakablealgorithm}
  {% \begin{breakablealgorithm}
   \begin{center}
     \refstepcounter{algorithm}% New algorithm
     \hrule height.8pt depth0pt \kern2pt % \@fs@pre for \@fs@ruled
     \renewcommand{\caption}[2][\relax]{% Make a new \caption
       {\raggedright\textbf{\ALG@name~\thealgorithm} ##2\par}%
       \ifx\relax##1\relax % #1 is \relax
         \addcontentsline{loa}{algorithm}{\protect\numberline{\thealgorithm}##2}%
       \else % #1 is not \relax
         \addcontentsline{loa}{algorithm}{\protect\numberline{\thealgorithm}##1}%
       \fi
       \kern2pt\hrule\kern2pt
     }
  }{% \end{breakablealgorithm}
     \kern2pt\hrule\relax% \@fs@post for \@fs@ruled
    \end{center}
  }
\makeatother

\begin{document}


\title{Scalable Breadth-First Search on a GPU Cluster}

% \author{\IEEEauthorblockN{Michael Shell}
% \IEEEauthorblockA{School of Electrical and\\Computer Engineering\\
% Georgia Institute of Technology\\
% Atlanta, Georgia 30332--0250\\
% Email: http://www.michaelshell.org/contact.html}
% \and
% \IEEEauthorblockN{Homer Simpson}
% \IEEEauthorblockA{Twentieth Century Fox\\
% Email: homer@thesimpsons.com}
% \and
% \IEEEauthorblockN{James Kirk\\ and Montgomery Scott}
% \IEEEauthorblockA{Starfleet Academy\\
% San Francisco, California 96678--2391\\
% Telephone: (800) 555--1212\\
\author{
\IEEEauthorblockN{Yuechao Pan$^{1,2}$, Roger Pearce$^2$, and John D.
Owens$^1$}
\IEEEauthorblockA{$^1$University of California, Davis; $^2$Lawrence Livermore National Laboratory\\
Email: ychpan@ucdavis.edu, rpearce@llnl.gov, jowens@ucdavis.edu}
}

\maketitle

\begin{abstract}

  \label{sec:abstract}
  On a GPU cluster, the ratio of high computing power to communication
bandwidth makes scaling breadth-first search (BFS) on a scale-free graph
extremely challenging. By separating high and low out-degree vertices,
we present an implementation with scalable computation and a model for
scalable communication for BFS and direction-optimized BFS\@. Our
communication model uses global reduction for high-degree vertices,
and point-to-point transmission for low-degree vertices. Leveraging
the characteristics of degree separation, we reduce the graph size to
one third of the conventional edge list representation. With several
other optimizations, we observe linear weak scaling as we increase the
number of GPUs, and achieve 259.8 GTEPS on a scale-33 Graph500 RMAT graph with
124 GPUs on the latest CORAL early access system.

\end{abstract}


\section{Introduction}
\label{sec:intro}
\John{I see this paper as ``how do we scale BFS on distributed GPU
  clusters''. The perfect audience is someone that wants to understand
  the design decisions that allow us to do that. It is very important
  at the beginning of this paper to explicitly say (a) how we want to
  be evaluated (in comparison to what kind of other systems, e.g.,
  only distributed-GPU) and (b) what we can do that was not possible
  before on whichever systems we pick. Is what-we-can-do ``scale to
  larger graphs than were previously possible''? Is it ``run faster
  than previous work on large graphs''? Both? This needs to be
  explicit.}

\YC{For (a) there are two comparisons we can make: for the same number
  of processors on more than 1 node, this work has the best
  performance; for the same graph scale, this work uses significantly
  less processors, without loosing too much performance, except for
  the latest work by Yasui (scale 33 with 128 CPU / 124 GPUs, our work
  has 1.5x performance). For (b) it is designing an implementation
  suitable for smaller number but fatter nodes, and that leads back to
  (a).}

% RAP:  Too many 'its' 'it', I tried to edit some out...
Breath-First Search (BFS) on graphs is a fundamental and important
problem that draws attention from a wide range of research
communities. It is a building block of more advanced algorithms that
involve graph traversals, such as betweenness centrality and community
detection. Traversals can be highly parallelizable; however, achieving good performance is
challenging, especially on scale-free graphs with wide ranges of degree
distribution.  This is due in part to low arithmetic computation density and
irregular memory access patterns caused by the algorithm and the graph
topology.  When running on distributed memory systems, high
communication cost adds additional challenges to achieve good
performance.  Because of the importance and challenging nature of
BFS at large scale, the High Performance Computing (HPC) graph community chose BFS as the first
benchmark in the Graph500~\cite{Graph500:2017}.  In addition to testing
hardware capability of HPC machines, the Graph500 has been a catalyst
for a series of algorithmic innovations~\cite{Buluc:2008:OTR,
Agarwal:2010:SGE, Beamer:2012:DBS} for HPC graph analytics.
\Roger{YC, can you dump a few cites for Graph500 papers here.   Frabrizio, Aydin, etc.}

Graphics Processing Units (GPUs) provide more computing power and
memory bandwidth than CPUs, and thus may be a good candidate for
a high-performance BFS\@. A fast BFS on GPUs is a challenge, however; irregular memory
access patterns and the workload imbalance caused by widely different
neighbor list lengths require optimizations to utilize the GPU
hardware. Another challenge is the low
per-processor memory size of the GPU (16~GB for the largest NVIDIA
GPUs), much smaller than the CPU's. Processing graphs larger
than one GPU's memory requires multiple GPUs and a distributed-memory
implementation.

On the algorithms side, Beamer, Asanovi\'{c} and Patterson~\cite{Beamer:2012:DBS} introduced
Direction-Optimizing (DO) BFS that significantly reduces traversal
workload on power-law graphs, such as those used by
Graph500 and social-network graphs. DOBFS's workload reduction
exacerbates the imbalance between highly efficient local GPU
computation and the relatively limited communication bandwidth in and
out of GPUs: a DOBFS implemented across multiple GPUs using existing
techniques will almost surely be limited completely by communication
bandwidth and will fail to scale. Our previous
work~\cite{Pan:2017:MGA:nourl} shows DOBFS is the most challenging algorithm
(among the five we tried) to scale even on multiple GPUs connected by
a PCI Express bus. Targeting a multi-node GPU cluster, with its
lower inter-node bandwidth, will be even more difficult. Existing work
on GPU clusters does not target DOBFS because of these challenges.

Our work targets the growing trend of multiple GPUs per compute node
on HPC systems.  CORAL/Sierra~\cite{CORAL} will be Lawrence Livermore National Lab's (LLNL)
newest supercomputer. This system will contain only a few thousand
compute nodes, compared to 10$\times$ that amount in previous supercomputers.
However, each node will feature more local computing power, mainly
from four Volta GPUs, and more memory. This change further
raises the computing power vs.\ communication bandwidth ratio. From a
BFS perspective, the graph partition on each GPU will be larger, while
the communication bandwidth for each GPU may not increase.
Thus, the available bandwidth per unit graph size decreases
significantly, and makes scaling on such systems harder.

In short, the challenges of a scalable (DO)BFS on GPU clusters are:
1) limited GPU memory---small per-GPU graphs will not be sufficient to
utilize the computing power of latest GPUs; 2) irregular memory access
patterns and unbalanced workloads, which together limit local traversal
performance; and 3) a high-computing-power to limited-communication-bandwidth ratio,
making scaling difficult.

Our work in this paper
%\footnote{This work was performed, in part, under the auspices of the U.S.
%Department of Energy by Lawrence Livermore National Laboratory
%under Contract DE-AC52-07NA27344.  Experiments were performed at the
%Livermore Computing facility.}
 targets an scalable implementation of (DO)BFS
for the CORAL early access system at LLNL called
\emph{Ray}~\cite{Ray} that can utilize the latest
hardware. Our implementation makes no CORAL-specific optimizations but
instead aims for generality to address any GPU cluster. We achieve
scalable performance up to a scale-33 RMAT graph on this machine. The
key idea allowing us to achieve scalable performance is that by
separating high- and low-degree vertices~\cite{Pearce:2014:FPT}, we
design and implement a scalable computation and communication model
that, for the first time, achieves scalable DOBFS on GPU clusters.
We make the following contributions:

\begin{itemize}
    \item Scalable BFS and DOBFS traversal results, reaching 260 Giga Traversal Edges Per Second (GTEPS) on
a scale-33 Graph500 RMAT graph with 124 GPUs, which is 18.5$\times$ better
weak-scaling performance than the best known GPU cluster work on the Graph500 list~\cite{Graph500:2017};
  \item An efficient graph representation that uses about half the
  memory as the conventional compressed sparse row (CSR) format;
  \item Fast and scalable local traversal on GPUs;
  \item A scalable communication model for DOBFS;
   and
  \item Several design decisions that may be useful for other programmers on
  similar systems.
\end{itemize}

\section{Related Work}
\label{sec:background}
\John{Probably want a DOBFS vs.\ BFS discussion, not just that ``DOBFS
  is faster'' but also that it's more interesting / more challenging /
  exercises more computation/communication patterns than just BFS\@.}
\YC{In Section~\ref{sec:DOBFS}}

\begin{figure*}
  \centering
  \begin{tabular}{b{0.49\textwidth}b{0.49\textwidth}}
    \includegraphics[width=0.49\textwidth]{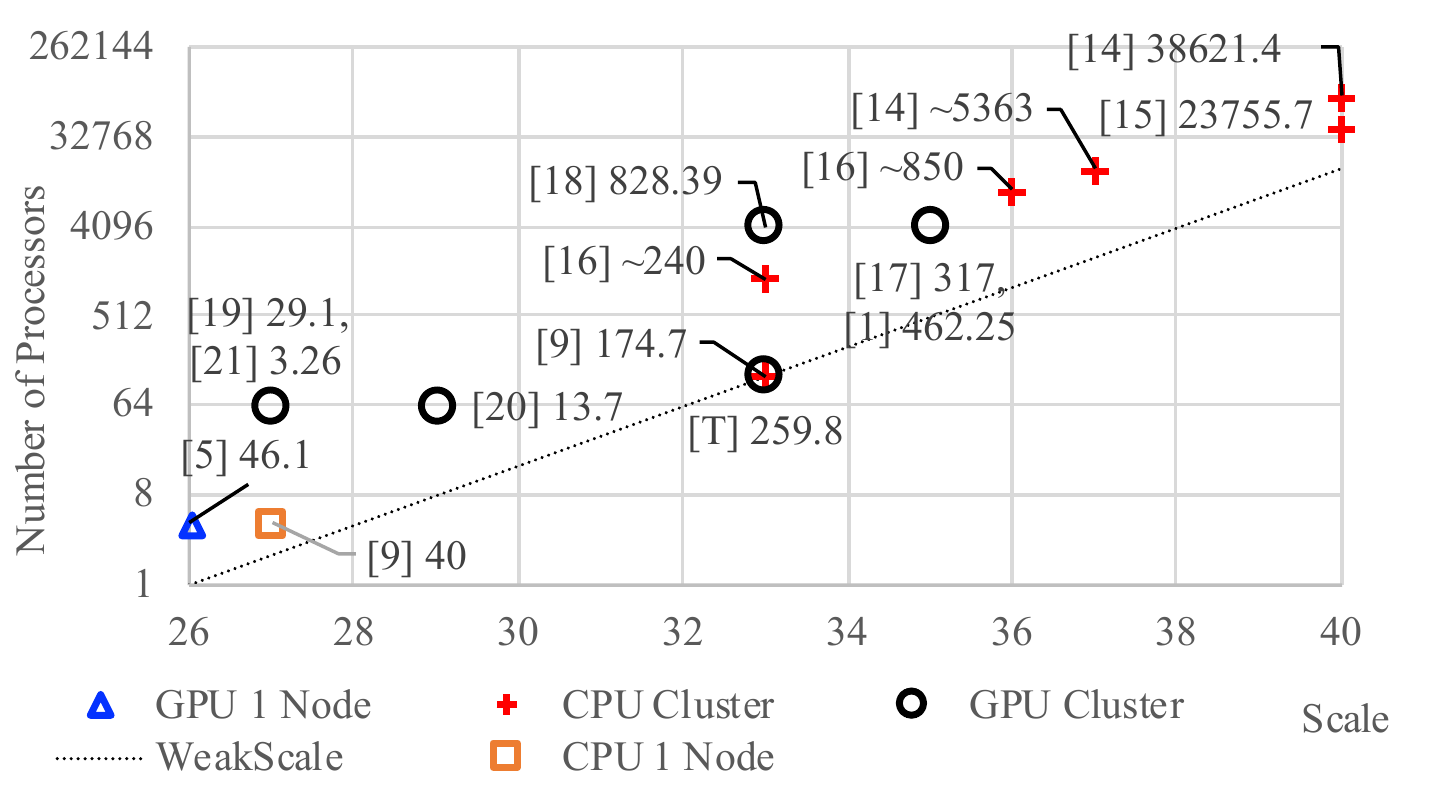}
    &
      \includegraphics[width=0.49\textwidth]{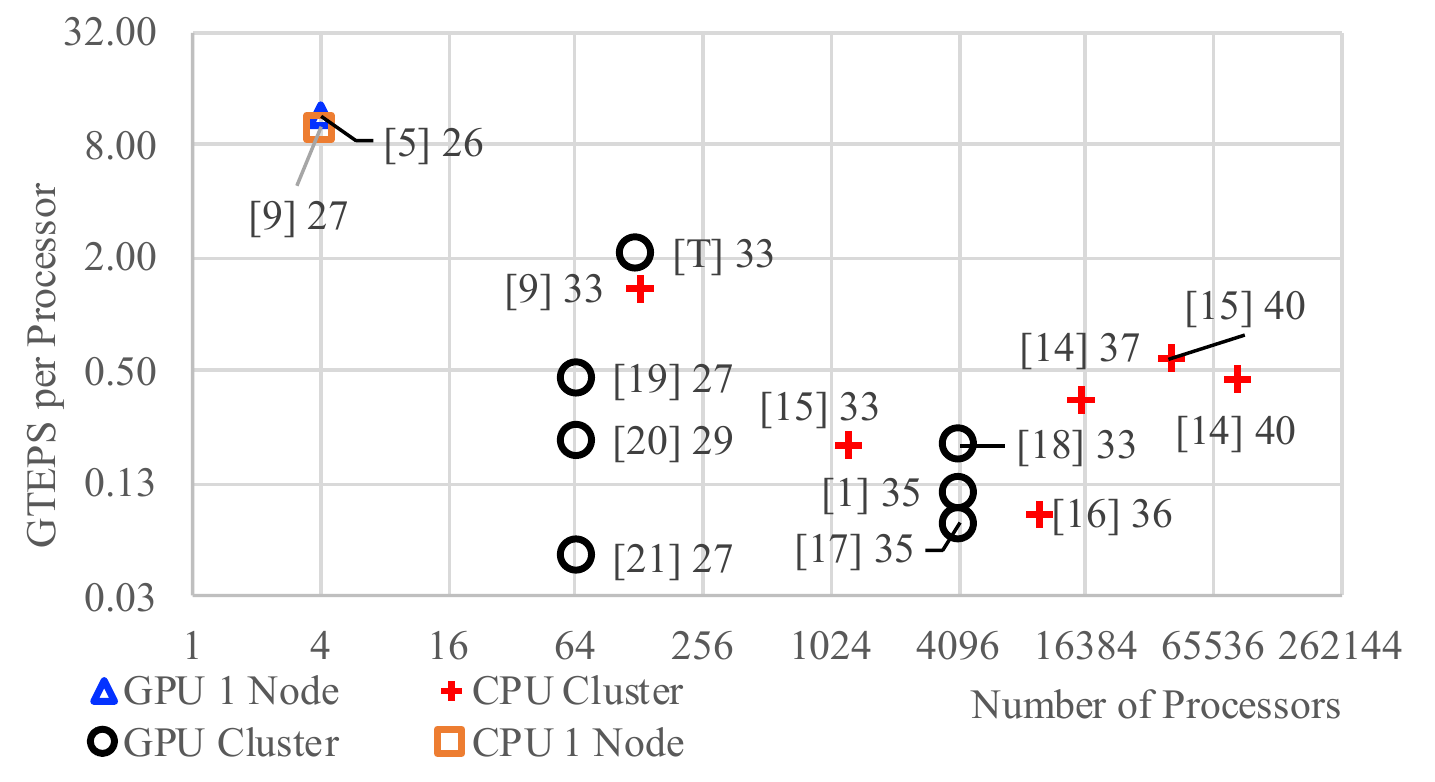}
  \end{tabular}
  \centering
  \caption{Placing our work (marked [T]) in the context of other
    large-scale BFS projects. GPU clusters are black circles and CPU
    clusters are red crosses. Two symbols mark top single-node
    CPU~\cite{Yasui:2017:FSE} and GPU~\cite{Pan:2017:MGA:nourl}
    accomplishments.
    \textbf{Left}: RMAT scale (graph size) vs.\ number of processors
    to process a graph at that scale. Results nearer the bottom right
    can process larger graphs with fewer processors. The dashed line
    represents the weak scaling line corresponding to our
    scale-processor count. Annotations mark aggregate GTEPS\@.
    \textbf{Right}: Cluster size vs.\ throughput (edges processed per
    second) per processor. Results nearer the top right sustain higher
    throughput with more processors. Annotations mark maximum RMAT
    scale.\label{fig:relatedworks}}

\end{figure*}

\subsection {Terminology}
\label{sec:terminology}
For ease of discussion, we define the following terms, and use them
in later sections of the paper.

\textbf{graph} A graph $G(V, E)$ is defined by its vertices $V$ and
edges $E$\@. In this paper, to study DOBFS scalability without
doubling the storage size, we assume the graph is
symmetric.

\textbf{$\boldsymbol{n}$} $= |V|$, the number of vertices in the graph.

\textbf{$\boldsymbol{m}$} $= |E|$, the number of edges in the graph.

\textbf{$\boldsymbol{p_\textit{rank}}$} the number of Message Passing Interface (MPI) ranks.

\textbf{$\boldsymbol{p_\textit{gpu}}$} the number of GPUs per MPI rank.

\textbf{$\boldsymbol{p}$} $= p_\textit{rank} \cdot p_\textit{gpu}$, the
number of GPUs used.

\textbf{$\boldsymbol{g}$} the inverse of inter-node communication
bandwidth.

\textbf{$\boldsymbol{\TH}$} the degree separation threshold (%details in
section~\ref{sec:separation}).

\textbf{delegates} vertices with out-degree larger than \TH\@.

\textbf{normal vertices} vertices with out-degree at most \TH\@.

\textbf{$\boldsymbol{E_{nn}, E_{nd}, E_{dn}, E_{dd}}$} normal
$\rightarrow$ normal, normal $\rightarrow$ delegate, delegate
$\rightarrow$ normal, and delegate $\rightarrow$ delegate edges.

\textbf{$\boldsymbol{d}$} the number of delegates in the graph.

\textbf{$\boldsymbol{S}$} the number of iterations (i.e., super-steps)
of running BFS on the graph, bounded by the diameter of the graph.

\subsection {Challenges with Scaling Directional Optimization}
\label{sec:challenges}

Directional optimization is a widely adopted optimization used in
high-performance BFS implementations. First described by Beamer,
Asanovi\'{c}, and Patterson~\cite{Beamer:2012:DBS},
it switches from the conventional
forward-push (i.e. top-down) direction to the backward-pull
(i.e. bottom-up) direction when the
workload of visiting all neighbors of the newly discovered vertices
from the previous iteration is greater than trying to find only one
previously visited parent of the unvisited vertices. The workload
savings from skipping a vertex's parent list once a valid one is found
can be huge, and it is very efficient for graphs with small diameters
and dense cores, for example, social networks and RMAT graphs.

\John{Don't know the citations, but we should cite 1D and 2D partitioning.}
\YC{I never saw reference to 1D partitioning, and 2D partitioning is from
a 2005 paper.}

However, conventional DOBFS implementations face scaling issues in a
cluster environment. When running in the backward-pull direction, each
active (unvisited) vertex must know the status of all its possible
parents. This information comes with a high communication cost. If the
graph is 1D-partitioned, it forces broadcasting the newly visited
vertices to all the peers that host their neighbors. In practice, this
often results in broadcasting the newly visited vertices to
\emph{every} peer, which is $8m$ bytes in communication volume, and
$8m/p \cdot g$ in communication time. If the graph is
2D-partitioned~\cite{Vastenhouw:2005:TDD},
it takes 2 hops to propagate the visiting status of vertices: one
reduction across the row direction, and one broadcast across the
column direction.

Let us use $n_t$ to indicate the number of vertices visited
in the forward-push iterations, and $S_b$ to indicate the number of
iterations in the backward-pull direction. We make the following assumptions: 1) row-
and column-wise vertex numbers are 32-bit; 2) reduction and broadcast works
in a tree-like manner, which gives
$\log{\sqrt{p}}$ communication for each column or row; 3) the same vertex is never visited in more than one
iteration, otherwise the communication cost will be higher; and 4) the processor grid is square, i.e., there are equal divisions
in the row and the column directions.
Then the total communication volume
for the forward direction is $8n_t\sqrt{p}\log{\sqrt{p}}$ bytes,
and it is $2n S_b \sqrt{p}(\log{\sqrt{p}}) / 8$ bytes for the backward
direction using compressed bit masks. The communication time is
$(4n_t + nS_b/8)((\log{\sqrt{p}})/ \sqrt{p}) \cdot g$.
%\footnote{This communication
%  result assumes 32-bit row- and column-wise vertex numbers, and that
%  reduction and broadcast works in a tree-like manner, which gives the
%  $\log{\sqrt{p}}$ communication for each column or row. It also
%  assumes that the same vertex is never visited in more than one
%  iteration, otherwise the communication cost will be higher. We also
%  assume the processor grid is square, i.e. three are equal divisions
%  in the row and the column directions.}
When the graph size and the number of nodes increase at the same rate
(weak scaling), the above communication cost will increase as $\sqrt{p}$, and this limits the scalability on large systems.
There are also increases in the computation workload:
instead of finding only one valid parent for each unvisited vertex,
the 2D partitioned case tries to finds $\sqrt{p}$ valid parents, one
in each of the $\sqrt{p}$ row-partitions of an unvisited vertex. When running
on large clusters, i.e. $\sqrt{p}$ is large, this workload increase
defeats the workload saving purpose of DO\@. In summary, both 1D
and 2D partitioning within a cluster on a DOBFS
present significant scalability challenges. \John{So this basically
  sounds like ``current partitioning algorithms are bad''.} \YC{Pretty much
  that, but almost everyone is still using them.}

Previous work on large-scale BFS falls into three categories.
\John{Which of these are DOBFS? Generally, all the big CPU ones? Do any
  of them do normal-delegate partitioning?}
\YC{Yes, all the CPU ones are DO, and Krajecki + all single node GPU onces;
non of them does normal-delegate partitioning. All cluster ones use
2D partitioning.}
Single-node projects,
either CPU or GPU, generally sustain the highest throughput per
processor but are limited by storage or compute to
relatively modest graph scales. The largest CPU clusters (tens of
thousands of nodes) have addressed the largest graph scales
($\geq 36$), whereas smaller-sized GPU clusters (thousands of nodes)
have not yet reached that scale. As a gross generalization, CPU
implementations are limited in scalability by computation (they must
add nodes to have more compute resources to process larger
graphs), whereas the GPU ones are limited by memory size (they
must add nodes to have more memory to store larger graphs).
We summarize this work in figure~\ref{fig:relatedworks}.

\subsection{BFS within Single Node}

Using GPUs in the same node for BFS yields impressive per-node
performance~\cite{Pan:2017:MGA:nourl, Yasui:2017:FSE, Liu:2015:EBG,
Ben-Nun:2017:GAA}, but
because all their communication is within a node and thus faster than
within a cluster, their per-node performance is superior to
cluster-based solutions. However, their graphs must fit into one
node's memory (GPU or CPU), and this inherently limits the maximum
size of a processed graph.

To break this memory limitation, other researchers have used a shared
memory architecture~\cite{Yasui:2017:FSE} or high-speed local
storage~\cite{Maass:2017:MPT}. The
shared memory architecture is essentially multiple nodes with unified
memory space, and it is less common \John{Adoptive? I don't know
  this word.} \YC{I mean very few multi-node system uses it.}
than distributed memory architectures. Using fast local
storage can help to process huge graphs with limited hardware
resources, but moving large amounts of graph data limits overall
traversal performance.

\subsection{BFS on CPU Clusters}

The Graph500 list is mostly CPU cluster
implementations~\cite{Ueno:2016:ESB, Lin:2017:SGT,
  Buluc:2017:DMB}, which use a large number of processors, typically
more than 10k, to reach the reported performance.

These implementations tend to use very specific graph representations~\cite{Buluc:2008:OTR, Ueno:2016:ESB}, which may not be GPU-friendly,
because their
complex memory access patterns bring extra irregularity and more
branching conditions, both of which
reduce achievable parallelism on GPUs. We instead choose a
standard graph representation (CSR). We expect our BFS implementation
will be used as a component of a complex workflow with many components
that use standard formats for passing data between them. Using
non-standard graph representations requires such a workflow to incur
an additional cost of format conversion, to duplicate graphs, or to
redesign other components, none of which are
desirable.

These implementations also generally use 2D partitioning to distribute
the graph across processors. 2D partitioning may introduce a high
communication cost (Section~\ref{sec:challenges}). As subgraph sizes on
each processor increase (to make full usage of more capable nodes as
the number of nodes decreases), the data transmitted per node will
increase together with the graph size, but the bisection network
bandwidth will be lower as the network shrinks. Machine-specific
network optimization could help, but this direction may make the
implementation less applicable to other systems.

The recent implementation by Yasui and Fujisawa~\cite{Yasui:2017:FSE} shows
a significant improvement in per-processor performance, using a shared
memory system with 128 processors. In this work,
the subgraph size on each processor is considerably larger than
previous BFS work on CPU clusters. With upcoming supercomputers
featuring a smaller number of nodes with more resources per node,
using larger sub-graphs per processor
may be more suitable for upcoming machines.

\subsection{BFS on GPU Clusters}

BFS on GPU clusters is a relatively recent topic of
study~\cite{Ueno:2013:PDB, Bernaschi:2015:EGD, Graph500:2017}
(citation~\cite{Graph500:2017} here refers to TSUBAME 2.0's number
31 ranking in the June 2017 Graph500 list. The achieved performance is
462.25~GTEPS with scale 35 using 4096 Tesla GPUs in 1366 nodes.
We can't find published work that references this particular record.), with
some recent work focusing on a smaller number of
GPUs~\cite{Fu:2014:PBF, Krajecki:2016:BTM, Young:2016:OCF}. None of
this work demonstrates competitive per-node performance vs.\ single-node
work, and none shows the combination of scalability and
performance per node that we demonstrate in this work.

\YC{
The citation numbers in Figure 1 is WRONG, will need update when
  all citations are checked in. The correct citation numbers are ([num
  in fig]-[actual ref num]) : [17]-~\cite{Ueno:2013:PDB},
  [19]-~\cite{Fu:2014:PBF}, % [10]-~\cite{Bisson:2015:PDB},
  [18]-~\cite{Bernaschi:2015:EGD}, [20]-~\cite{Krajecki:2016:BTM},
  [21]-~\cite{Young:2016:OCF}, [16]-~\cite{Buluc:2017:DMB},
  [14]-~\cite{Ueno:2016:ESB}, % [13]-~\cite{Ueno:2017:EBF},
  [9]-~\cite{Yasui:2017:FSE}, [15]-~\cite{Lin:2017:SGT},
  [5]-~\cite{Pan:2017:MGA:nourl}, [1]-~\cite{Graph500:2017}}

\section{Graph Representation}
\label{sec:graph}
\John{So before this section, I'm thinking maybe having a ``Baseline
  Implementation'' section. What is the straightforward way to
  implement BFS on a GPU cluster without any of your cool ideas (this
  should be roughly equivalent to existing work in the area)? First,
  describe what that system looks like. Then describe what its
  limitations are. Then \emph{this} section can be where you describe
  your ideas and, qualitatively, why they're better. (Then in results,
  you can quantitatively show they're better.}
\YC{Added a subsection
  in background, talking about DOBFS}
\John{Very important: How do the
  ideas in this section address the DO scalability challenges from the
  previous section?}
\YC{This section only address the potential graph storage size issue,
  with some analysis and comparison at the end of section.}

The key to a scalable DOBFS on a GPU cluster is to (a) maximize
the fraction of the graph that can be stored on one GPU, thus allowing
fast computation with no communication on that portion of the graph,
and (b) optimize the communication between GPUs, which would otherwise
limit scalability, even within a single
node~\cite{Pan:2017:MGA:nourl}.

\subsection{Separation of Vertices}
\label{sec:separation}

Our design to accomplish these goals starts from a simple but powerful
idea that we have pursued in previous work on CPU
clusters~\cite{Pearce:2014:FPT}: separate the vertices into two sets
by out-degrees, and treat them differently. The separation point
between the two, the threshold out-degree \TH, is an important tuning
parameter, and we will show how it affects the overall performance in
upcoming sections. We call vertices with more than \TH\ direct
neighbors the \textbf{delegates}, and the rest \textbf{normal
  vertices}.

The intuition behind this design choice is that in local traversal,
vertices at different ends of the degree distribution should have
different load-balancing strategies; and in communication,
vertices that almost every GPU touches should not be treated the same
as those needed by very few GPUs. By separating vertices into
different sets, we can pursue different strategies in graph
representation, local traversal, and communication on those
sets, which we describe below.

% \John{Why? What is the intuition behind this separation?} \YC{The
% main reason is high and
% low degree vertices need different treatments} \John{Again,
% \emph{why} do they need different treatments? You have to justify
% why. I think you want to say that \emph{some} edges have to cross
% between GPUs. Of course you want to minimize this, but it's
% inevitable. But if you have to pick edges to cross GPUs, you prefer
% edges attached to normal vertices instead of delegates, because
% \ldots} \YC{Added in more explanation and rearrange the paragraph}

\subsection{Distribution of Edges}
\label{sec:distribution}

\begin{breakablealgorithm}%[H]
  \caption{Edge Distributor \label{algo:distributor}}
\begin{algorithmic}[1]

  \Statex Let $P(v) = \bmod(v,p_\mathit{rank})$
  \Statex Let $G(v) = \bmod(v / p_\mathit{rank}, p_\mathit{gpu})$
\For {each edge $(u \rightarrow v)$}:
    \If {$u$ is normal}:
        to rank $P(u)$, GPU $G(u)$
    \ElsIf {$v$ is normal}:
        to rank $P(v)$, GPU $G(v)$
    \ElsIf {$(\mathit{OutDegree}(u) < \mathit{OutDegree}(v))$}:
        \State to rank $P(u)$, GPU $G(u)$
    \ElsIf {$(\mathit{OutDegree}(u) > \mathit{OutDegree}(v))$}:
        \State to rank $P(v)$, GPU $G(v)$
    \Else:
        to rank $P(\min(u, v))$, GPU $G(\min(u, v))$
    \EndIf
\EndFor
\end{algorithmic}
\end{breakablealgorithm}

On scale-free graphs, most storage is devoted to edges, not vertices.
We distribute edges to individual MPI ranks, and then to individual
GPUs within the same rank, using the distributor described in
Algorithm~\ref{algo:distributor}. In it, we divide edges into four
categories depending on the type of their source and destination
vertices (normal or delegate). Our edge distributor has the
following advantages:

\textbf{Simple} The location of an edge can be easily computed from
its index locally without table lookup or remote query.

\textbf{Symmetric} Except for normal to normal edges,
subgraphs on individual GPUs are symmetric. Because we make
an edge pair of opposite directions for an undirected edge,
they need to be on the same GPU to preserve the correctness
of DOBFS without a global traversal direction. Otherwise if
the traversal directions of the edge pair are opposite to their
respective directions, both edges will be ignored.
\John{I don't
  understand this previous sentence.} \YC{En, maybe an example is better?
  consider edge pair a $\rightarrow$ b and b $\rightarrow$ a,
  they are on different GPU and we need to go from a to b,
  if the traversal direction of a $\rightarrow$ b is backward, and that
  for b $\rightarrow$ a is forward, this edge pair will be ignored,
  which is an algorithmic bug.}

\textbf{Bounded size} The number of possible destination
vertices for non-(normal to normal) edges on each GPU are bounded:
the number of normal vertices is
at most $n / p$, and the number of delegates is at most $d$. Thus
vertex indices for these edges can be
represented as 32-bit numbers locally, and converted back to 64-bit
when necessary for communication. This allows us to store more of the
graph in a fixed-size memory.

\textbf{Balanced} This distribution prioritizes placement of vertices
with lower out-degrees. Neighbor lists of high-degree vertices are
distributed according to the destination vertices, and scattered
across the entire cluster. The number of edges in the partitioned
subgraphs on individual GPUs are very close to each other, giving each
GPU a balanced workload.

\begin{figure}
  \centering
  \begin{tabular}{c} %m{0.49\textwidth}} % m{0.49\textwidth}}
    \includegraphics[width=0.34\textwidth]{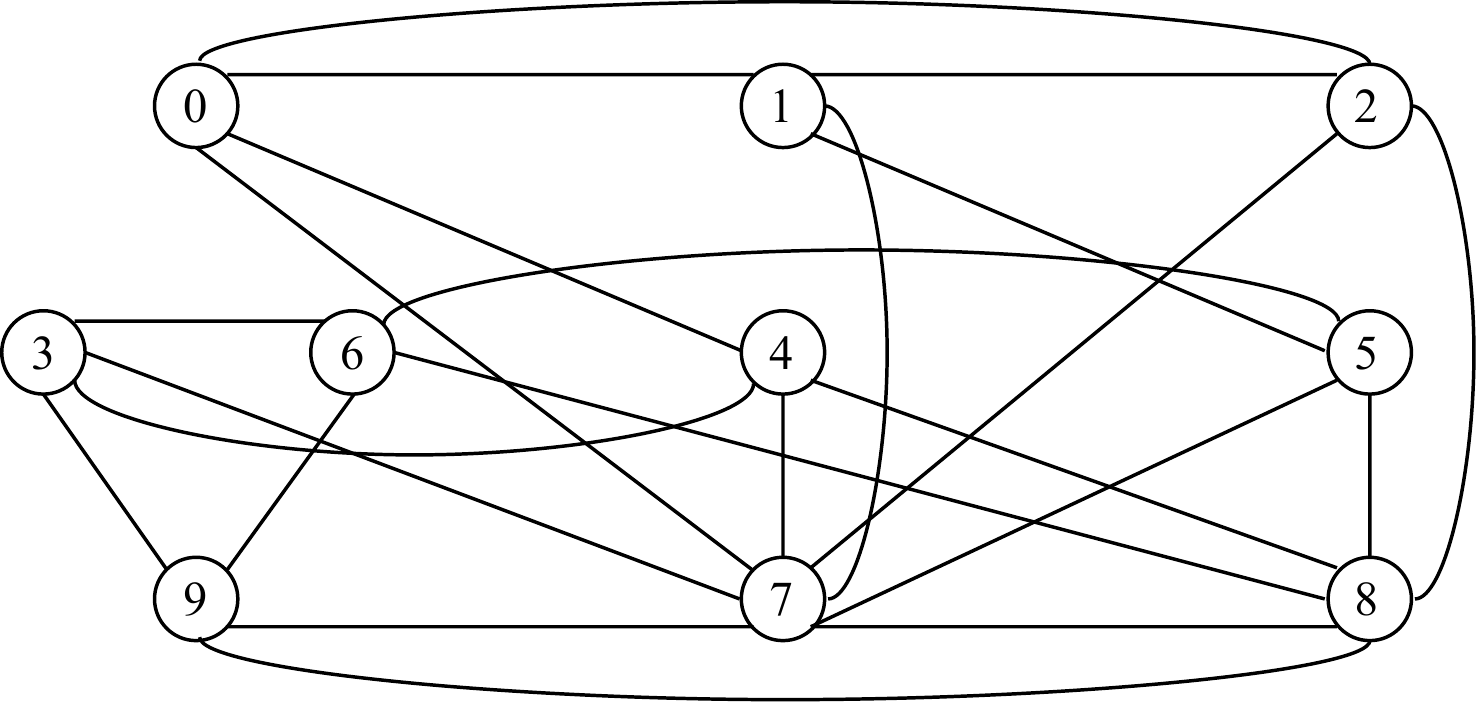} \\
    \includegraphics[width=0.37\textwidth]{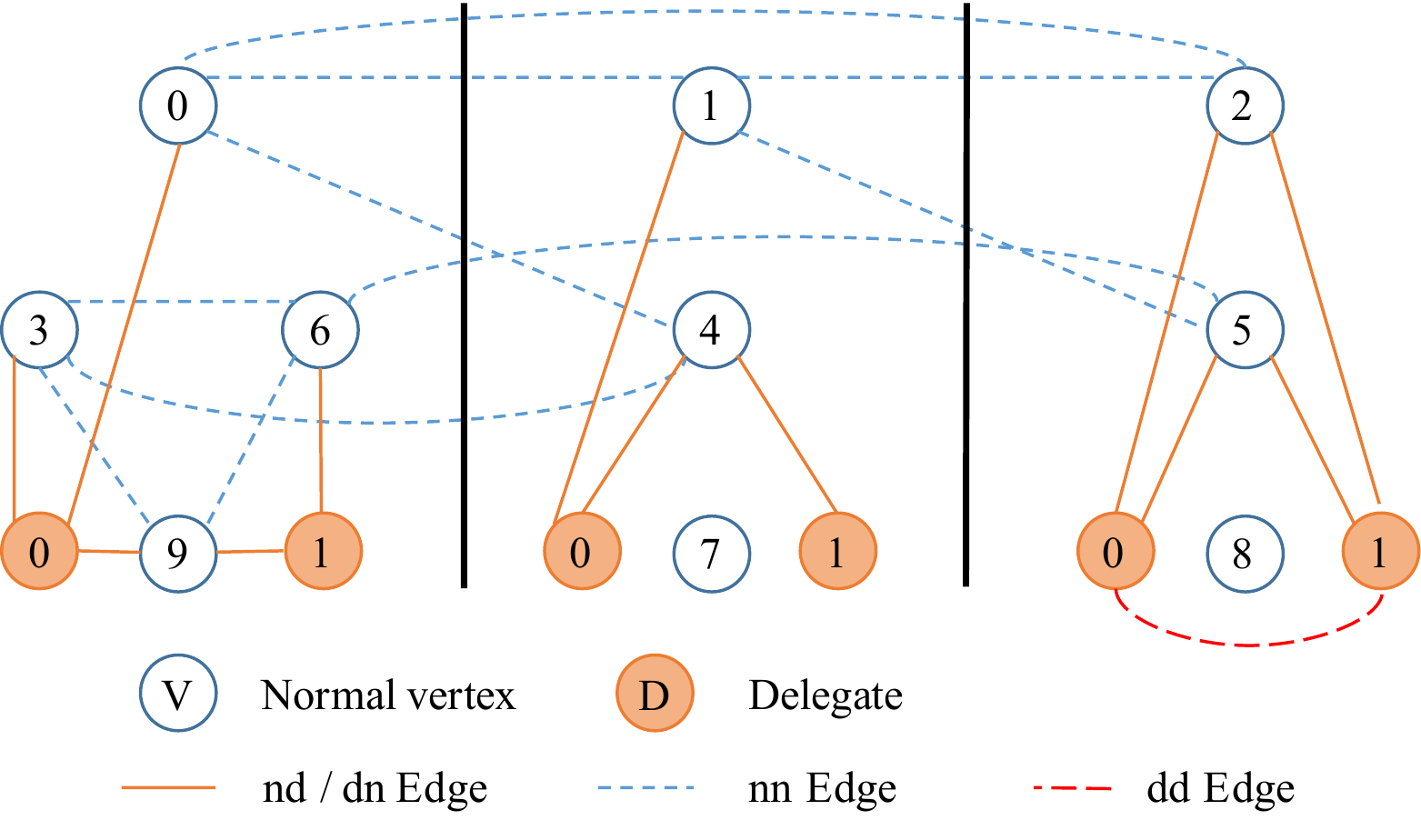}
  \end{tabular}
  \centering
  \caption{Example of degree separation and edge distribution for a
    graph with 3 partitions and degree threshold 5. Top: the original
    graph. Bottom: subgraphs after making vertex 7 as delegate 0, and
    vertex 8 as delegate 1. nd, dn, nn and dd refer to normal to
    delegate, delegate to normal, normal to normal, and delegate to
    delegate edges, respectively.
     \label{fig:partition} \John{Why are 7 and 8
    still assigned to a partition?} \YC{The normal vertex id for
    delegates are kept, but there is no edge connected to them. Global
    renumbering may remove these holes + all zero-degree vertices,
    but not yet done that. It may not be applicable to algorithms
    that required actual vertex ids.}}
\end{figure}

Figure~\ref{fig:partition} illustrates our vertex separation and edge
distribution strategies. Vertices 7 and 8 have out-degree more than
\TH, which is 5 in this example, and they are converted to delegates
0 and 1, respectively. All partitions keep a copy of the delegates,
and all edges involving the delegates are changed to the local copies.
After this operation, only edges between normal vertices require
communication across GPUs; all other edges are between two vertices on
the same GPU\@. This is the right choice because normal vertices are
the ones with the fewest neighbors and thus the least communication.
Any delegate-related communication is performed using
global reductions (details in
section~\ref{sec:communicate}). \John{What is the delegate-related
  communication here? Not clear at this point in the paper.}
\YC{I think a forward pointer here should be okey?}

\subsection{Efficient Graph Storage}

\begin{table}
  \centering
  \begin{tabular}{c c c}
  \toprule
  sub-graph & row offsets & column indices\\
  \midrule
  nn & $n / p \cdot 4$ & $ |E_\textit{nn}| / p \cdot 8$ \\
  nd & $n / p \cdot 4$ & $ |E_\textit{nd}| / p \cdot 4$ \\
  dn & $d \cdot 4$ & $ |E_\textit{dn}| / p \cdot 4$ \\
  dd & $d \cdot 4$ & $ |E_\textit{dd}| / p \cdot 4$ \\
  Total & $8n + 8d \cdot p$ & $4m + 4|E_\textit{nn}|$ \\
  \bottomrule
  \end{tabular}
  \caption{Memory usage for subgraphs in bytes.\label{tab:memory}}
 \end{table}

The bounded-size feature of our edge distributor is critical for
processing huge graphs within limited GPU device memory, and makes
processing larger graphs using the same number of GPUs possible. When
the number of local normal vertices and the number of delegates are
bounded by $n / p$, with the exception of destinations of nn edges, we
can use 32-bit values to store local normal vertex and global delegate
ids, instead of 64-bit values. This provides significant savings on the
memory storage footprint of the graph. As listed in
Table~\ref{tab:memory}, the total memory usage for all subgraphs on
the GPUs is $8 n + 8 d \cdot p + 4 m + 4|E_\textit{nn}|$ bytes. In
practice, when using suitable values of \TH\
(Section~\ref{sec:parameters}), while still using CSR format for each
sub-graphs, the above memory usage is only about
one third of the conventional edge list format ($16m$ bytes), and a
little more than half of CSR format ($8n + 8m$) without the degree
separation.

It is possible to utilize CPU memory and handle graphs larger than GPU
memory, with different techniques~\cite{Green:2016:CSD, Sakharnykh:2016:BGM}.
However,
the current latency and bandwidth differences between GPU memory and
the GPU-CPU connection would impose a high performance penalty. This
decision could be revisited when CORAL is fully equipped with NVLink2,
which doubles CPU-GPU bandwidth, in the near future. In
this paper, we only focus on graphs that fit in GPU memory.

\section{Local Computation}
\label{sec:computation}
\begin{figure}
  \centering
  \includegraphics[width=0.49\textwidth]{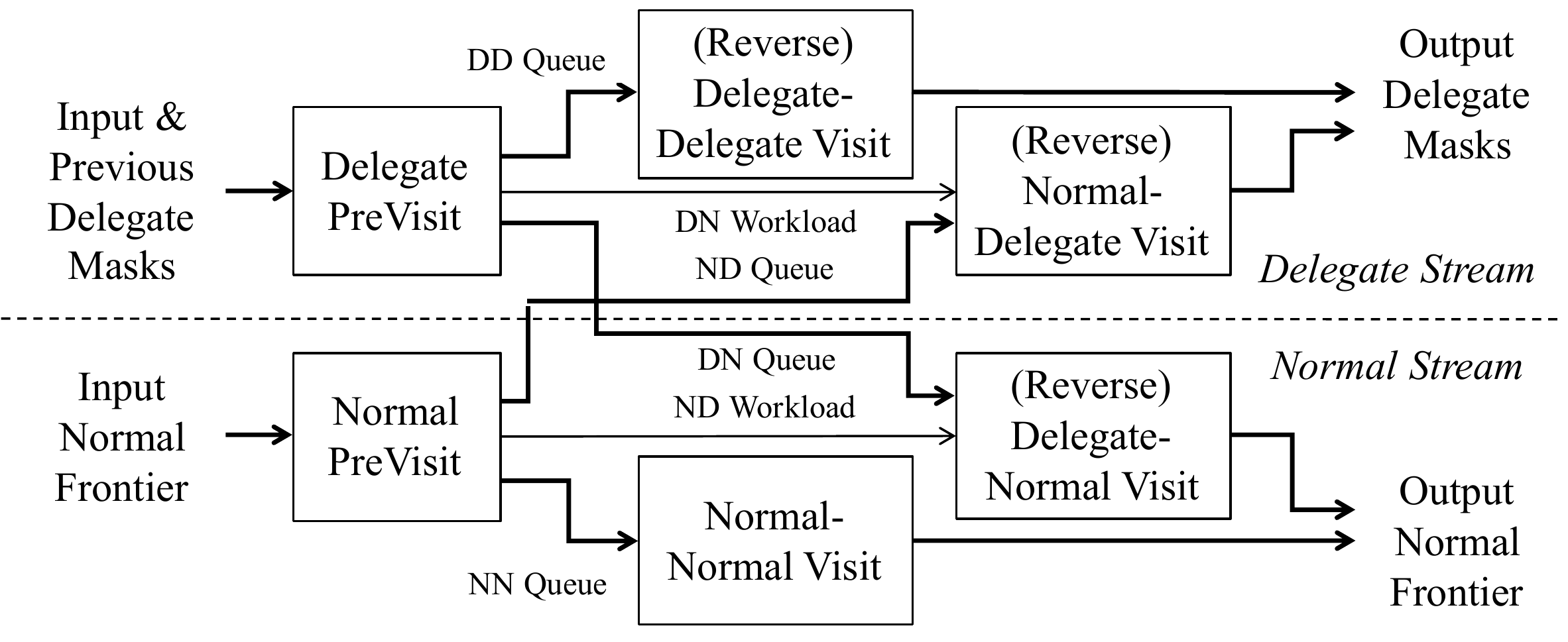}
  \centering
  \caption{Local computation for one BFS iteration,
    showing data dependency and stream allocation.
    \John{Correct ``Pervious'' to ``Previous''.} \YC{done}
    \label{fig:computation}}
\end{figure}

On each GPU, we now have 4 different subgraphs: normal to
normal (\textbf{nn}), normal to delegate (\textbf{nd}),
delegate to normal (\textbf{dn}) and delegate to
delegate (\textbf{dd}). While we could apply the exact same strategies
to each of them, we note that their different characteristics motivate
different load-balancing strategies for traversal
(Section~\ref{sec:traversal}), different direction switching
conditions for DOBFS (Section~\ref{sec:DOBFS}), and different input
from/output to the communication model
(Section~\ref{sec:communication}). Because subgraphs can be processed
in parallel, we can achieve some overlap between computation and
communication in our processing pipeline (Fig.~\ref{fig:computation}).

At a high level, we separate local traversal on the four subgraphs
into a delegate stream and a normal stream, depending on the
destination type of the edge, as two \textit{cudaStreams}. Each stream
begins with a ``previsit'' kernel, used to preprocess the inputs. This
includes marking level labels for input vertices,
filtering out duplicates and zero-out-degree vertices, forming the
queues of vertices to be visited by the visit kernels, and calculating
the would-be workload for these kernels, which is important for
the direction decisions in DO\@. Then each
stream spawns a ``visit'' kernel for the two edge types in the stream.
The two streams run independently of each other, except when
dependencies are established explicitly  (Fig.~\ref{fig:computation}).

\subsection{Forward Traversal}
\label{sec:traversal}
The visited status of delegates are maintained by bitmasks, with each
delegate only occupying 1 bit. This is an
effective way to store and communicate the status of high out-degree
vertices. We use advanced load balancing techniques for the visiting
kernels: the delegate to delegate visit kernel uses
merge-based workload partitioning~\cite{Davidson:2014:WPG:nourl}, because the dd
subgraph covers a wide range of degree distribution, and has large
average out-degrees; the other visit kernels use thread-warp-block
dynamic workload mapping~\cite{Merrill:2012:SGG}, based on the fact that the
out-degree range of dn, nd, and nn subgraphs are all limited, and the
average out-degrees are low. \John{Why does dn have a low out-degree?}
\YC{dn is the reverse of nd, and the number of delegates is more than
the number of local vertices, so the average of dn is lower than that
of nd, which is lower than \TH, normally much lower. }

\subsection{Directional Optimization}
\label{sec:DOBFS}

Not all subgraphs benefit from directional optimization. We do not use
DO for normal $\rightarrow$ normal visits, because the nn subgraph on
each GPU is not symmetric, the range of destination vertices of nn
edges are unbounded, and most importantly, DO is not efficient for the
very low in-degree nn subgraphs. Without separating the graph,
skipping the nn portion from using DO is impossible.

On each GPU, we keep a source list of the normal-to-delegate subgraph,
i.e., all the normal vertices that have edges
pointing to delegates. These are exactly the potential destination
vertices in the reverse subgraph, i.e., the delegate to
normal subgraph. When running in the backward-pull direction for a
delegate to normal visit, we use the normal-to-delegate
subgraph, and start from its source list. For the same
purpose, we keep source masks for the dd and dn subgraphs. Keeping
source lists and masks avoids vertices that may not find local
parents, and provides more accurate workload prediction.

The traversal direction is decided based on a workload comparison,
computed in each iteration, between the forward and the backward
directions. The forward workload \FV\ is calculated by the
previsit kernels as the sum of neighbor list lengths from the source
vertices to be visited. The backward workload \BV\ is calculated
using the estimated number of parents to check until finding the first
visited one. Let: % $U$ being unvisited sources in the reversed graph,
% $q$ the input frontier length, $s$ the number of unvisited sources in
% the forward graph, $a$ the probability that a potential parent is newly
% visited, and $od(u)$ the out degree of $u$; then:
\begin{align*}
    U &= \text{unvisited sources in the reversed graph};\\
    q &= \text{input frontier length};\\
    s &= \text{number of unvisited sources in the forward graph};\\
    a &= \text{probability that a potential parent is newly visited};\\
      &= q / (q + s);\\
    od(u) &= \text{out degree of u};\\
\text{Then: }
   BV &= \sum_{u \in U} ((1-a)^{od(u)}
         + \sum_{i=0}^{od(u)-1} a(1-a)^i)\\
    &= \sum_{u \in U} \frac{1-(1-a)^{od(u)}}{a} \approx |U|/a \\
    &\text{assuming }od(u)\text{ is large, \& }a\text{ not too small}\\
    &= |U|(q + s) / q
\end{align*}

Starting from the forward-push direction, with two direction-switching
factors
$\mathit{factor0}$ and $\mathit{factor1}$, the visiting direction is decided as:\\
\indent if current direction is forward, and $\FV > \mathit{factor0}\cdot \BV$\\
\indent \hspace{15pt} then switch to backward;\\
\indent if current direction is backward, and $\FV < \mathit{factor1}\cdot \BV$\\
\indent \hspace{15pt} then switch to forward;\\
\indent otherwise keep current direction.

\noindent
No matter which direction a visiting kernel takes, it only affects the
kernel itself, and the input and the output are the same. The three
visiting kernels with DO have three sets of direction-switching
factors. This allows the
kernels to switch for their own optimized conditions.

Our strategy for DOBFS results in a smaller workload than a 2D
partitioning strategy. In our strategy, for normal vertices, only one
GPU must do the reverse visiting for each individual normal vertices.
Only the delegates may need to have more than one GPUs visiting their
parents, and moreover, the delegates are only a small portion of all
vertices. Let $m'$ be the number of edges the DOBFS algorithm would
need to visit if the graph was traversed by a single processor. Then
the workload of our DOBFS implementation would be bounded by
$m' + dp\cdot b$, where $b$ is the average number of parents a
delegate must search on each GPU before finding a visited one.
While keeping $d$ in the order of $O(n/p)$, the term $dp\cdot b$ is
scalable even when $p$ is large, because it is in the order of $O(nb)$
and $b$ is not a large number---only delegates with
very large out-degrees are distributed across a large number of nodes, and
delegates with large out-degrees tend to be close to portions of the
graph with high connectivity, which reduces the number of neighbors to
try before finding a visited one.

\section{Communication}
\label{sec:communication}
\begin{figure}
  \centering
  \includegraphics[width=0.49\textwidth]{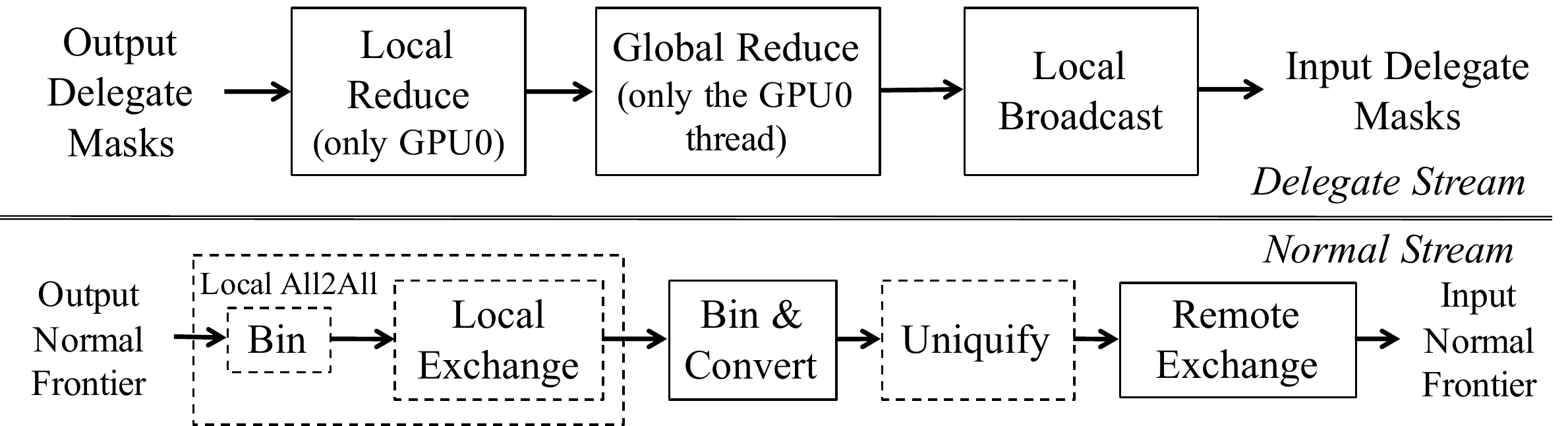}
  \centering
  \caption{Communication flow
     \label{fig:communication}}
\end{figure}

\label{sec:communicate}

Because local computation performance is increasing more quickly than
interconnect bandwidth, designing for scalable communication is more
important for graph processing than ever before. Our scalable
communication model (shown in Fig.~\ref{fig:communication}) adopts
different strategies for delegates and normal vertices.

\subsection{Communication for Delegates}

The visited status of delegates may be updated by any GPU, or consumed by any
GPU\@. We thus use a global reduction to gather and distribute the delegate
mask updates whenever any update occurs in a
iteration. The reduction is done in two phases: locally across peer GPUs, and
globally across different MPI ranks. During the local phase, all GPUs in the same
MPI rank push their updated masks to GPU$_0$, and GPU$_0$ performs the
reduction in parallel. During the global phase, only GPU$_0$ (more
accurately, the CPU thread that controls GPU$_0$) participates, and all GPUs in the
same MPI rank consume the resulted masks for the next iteration. We utilize fast
GPU-GPU data channels and the GPU's parallel computing capability for the local
phase, and efficient \textit{MPI\_(I)AllReduce} calls for the global phase.

The cost of this delegate communication is small. For each iteration that has
updates to the delegate masks, the communication volume is $2dp_\textit{ranks}/8$
bytes, and the communication time is $d \log{p_\textit{ranks}} /4 \cdot g$, assuming
the global reduction is done in a tree-like manner. The delegate reduction might
run on every iteration, which gives the total communication cost as
$d \log{p_\textit{rank}} /4 \cdot gS$. However, for graphs with more concentrated cores,
the delegate updates will finish faster than normal vertices, which reduces the
number of iterations that require delegate communication. In practice, we keep
$d$ (the number of delegates) low, so that the size of delegate masks, $d/8$, is
under the limit of several tens of MBs.

% For BFS, each delegate only needs 1 bit for the visited status,
% indicating whether that delegate been visited or not; as a result, we
% can use compact bit masks. For other graph algorithms that require
% more bits of state
% (i.e., ranking scores for PageRank), the communication volume for delegates
% will increase accordingly. Our delegate communication should still be scalable,
% except for computation that involves updates to the delegate status for too many
% iterations.

\subsection{Communication for Normal Vertices}

The basic communication model for normal vertices is point-to-point transmission
via \textit{MPI\_Isend} and \textit{MPI\_Irecv}. We use a non-blocking version
to keep the pipeline running and take advantage of possible workload overlaps.
The total communication volume is $4|E_\textit{nn}|$ bytes, assuming each nn edge is a
cutting edge (i.e., a edge with end points on two different GPUs). The
communication time is $4|E_\textit{nn}|/p \cdot g$. Note that only the outputs from
nn edge visits may result in direct remote normal vertex updates: the results
from dd and nd edge visits are communicated via global delegate mask reduction,
and the updates from dn visits are always local, as a result of our edge
distributor (Algorithm~\ref{algo:distributor}). When setting \TH,
the degree threshold, to an optimal value, the nn edges are only a small portion
of the graph, and the resulting normal communication is a lot less than
$m$.

The normal vertex exchange requires some extra local computation, such
as binning (group vertices need to be sent to the same GPU together)
and vertex number conversion (from the 64-bit global ids used in nn
edge destinations to 32-bit local ids at destination GPUs). These
computations are done on GPUs. The workload is in the order of
$O(|E_\textit{nn}|/p)$ on each GPU for all iterations combined. This is a small cost
compared to the traversal workload, and does not affect the
scalability of our BFS implementation.

We also tried two optimizations to reduce communication cost. The first one
is called \textit{Local All2all}: prior to the remote vertex exchange,
we first run a local
exchange to gather vertices going to GPU$_x$s in all MPI
ranks on the local GPU$_x$\@. As a result, normal vertex exchanges only occur among
GPU$_0$s, among GPU$_1$s, etc., but never between GPU$_0$s and GPU$_1$s, etc. This
reduces the number of communication pairs from $p^2$ to
$p^2 / p_\textit{gpu}$, each of which has more vertices to send.
In turn, this allows a second optimization,
uniquification, which removes duplicated vertices going to the same GPU\@.
However, because relatively few individual destination vertices of nn edges are on a
given GPU or node, with the expected value capped by $\TH / p_\textit{rank}$, the chance to find
duplications is small, and may not be sufficient to overcome the extra
computation. We show our findings in the next section.

Combining the communication for delegates and normal vertices together, we have
a model that has at most $dp_\textit{ranks}/4 \cdot S + 4|E_\textit{nn}|$ bytes
total volume and $(d \log{p_\textit{ranks}} /4 \cdot S + 4|E_\textit{nn}|/p)g$
communication cost. For graphs that have a small number of vertices covering
a large portion of edges, the number of iterations $S'$ that need delegate
masks exchange, is less than $S$; for the graphs we tested, $S'$ is about half
of $S$. With suitable values of \TH\ (Section~\ref{sec:parameters}), we saw
delegate mask reduction and normal vertex exchange taking roughly the same amount
of time. Under these conditions, we approximate our communication cost as
$d\log{p_\textit{rank}}/4 \cdot Sg$. We also keep $d$ on the same scale as $n / p$,
more accurately, under the value of $4n/p$ in practice. As a result, the communication
cost is $n \log{p_\textit{rank}} /p \cdot Sg$. It starts from $n\cdot Sg$ on
single node, and grows on the order of $\log{p_\textit{rank}}$ when $n$ and $m$
increase at the same rate as $p$ (weak scaling). This growth is slow, and more
scalable than the $\sqrt{p}$ growth order of conventional 2D partitioning methods.
Thus, we argue that our communication model is more scalable.

\section{Results}
\label{sec:results}

\John{I'd like to see a caption (without the graph) for every plot/table you will put in this section. The caption needs to say what the plot says (will say) and what the conclusion is that the reader will draw from this plot. It is OK if you write more in the caption than we will eventually use. Make sure it is very specific on the experiments that need to be run, what will actually be on the plot, and the conclusion that you will draw from it.}
\YC{All figures and tables are checked-in. Actually most of them come from
parameter sweeps}

\subsection{Testing Environment}

%\begin{figure}
%  \centering
%  \includegraphics[width=0.49\textwidth]{fig/p2p.pdf}
%  \centering
%  \caption{Communication throughput with 32 nodes by 4 threads per node,
%  when changing MPI message sizes. Grouped by data size per $<$source, destination$>$ thread pair.
%     \label{fig:p2p}}
%\end{figure}

\subsubsection{Hardware}

Our implementation targets an early access system (\emph{Ray}) of LLNL's upcoming
CORAL/Sierra supercomputer. The current system has more than 40
compute nodes; each features two
10-core IBM Power8+ CPUs at 2.06~GHz with 256~GB CPU memory. Each CPU has two
NVIDIA Tesla P100 GPUs; the two GPUs and the CPU are connected by high-speed
NVLink~\cite{NVIDIA:2014:NNH} with 40~GB/s bandwidth in each direction. % Each GPU has
% 56 multiprocessors, and 64 CUDA cores per multiprocessor at 1.48 GHz, and 16 GB
% memory with 732 GBps bandwidth.
Each socket has a Enhanced Data Rate (EDR) 100~Gbps InfiniBand
connection to a network with FatTree topology.

Because the interconnection speed is higher than a conventional
cluster, and because GPUs achieve their best performance only when
fully occupied by a sufficient workload, we first test how the network
performs with different message sizes.
%Figure~\ref{fig:p2p} shows the
% \textit{MPI\_Isend / MPI\_Irecv} throughput as a function of MPI
% message sizes.
In an experiment, we use 32 nodes, each with one MPI rank,
and 4 CPU threads, and each thread sending MB-sized data to all threads on
other nodes, to simulate a scenario where each of the 128 GPUs
sends out data to the 124 GPUs on other nodes. After sweeping through the message
size from 128~kB to 16~MB, we found that the optimal message
size is about 4~MB for data larger than 2~MB\@. While this is much
larger than normal MPI usage, it is the best fit for the GPUs. With
smaller data (under 2~MB), the network appears to do a better job with
caching, and the differences between message sizes are not that
significant. \YC{should make this paragraph shorter}

\subsubsection{Software}

The cluster runs on 64-bit Linux with GPU driver version 384.59. The
compilation toolchain includes gcc 4.9.3, cuda 9.0.167, and
spectrum-mpi 2017.08.24 with OpenMP support. \textit{nvcc} options are
\textit{-O3 --std=c++11 --expt-extended-lambda}. The GPU target is set
to the hardware's SM version (i.e., 6.0 for P100 GPUs).

\emph{At the time of our experiments}, we faced the following current limitations
with \emph{Ray}. Many of these will be addressed in the
full system or with system software updates.

\begin{itemize}
  \item Network Interface Controller (NIC)-GPU Remote Direct Memory Access (RDMA)
  was not planned for \emph{Ray}; instead all NIC-GPU traffic goes through CPU memory.

  \item Asynchronous GPU memory copy was not supported by the MPI implementation;
  as a workaround, our implementation copies data from GPU memory to CPU memory
  with appropriate \textit{cudaMemcpyAsync} calls, then issues MPI calls from the
  CPU memory, and copies the data from CPU to GPU on the receiving end.

  \item Random delays of $\sim$100 ms were observed
  when consuming data on CPU right after receiving them from
  unblocking MPI calls; as a result, we only use the CPUs for GPU workload
  scheduling and data movement controls in our experiments.

  \item Degraded data movement performance between CPUs and GPUs were observed
  on some nodes; all but one experiments only use up to 124 GPUs on 31 selected
  nodes to avoid this issue; the experiment with the WDC 2012 graph includes
  3 GPUs affected by this issue, out of 160 GPUs on 40 nodes.

\end{itemize}

\subsubsection{Reporting}

We use RMAT graphs for testing our BFS implementation. The RMAT graph
generator is a distributed GPU implementation conforming to the Graph500
specifications~\cite{Graph500:2017}. The edge factor is 16, and the RMAT parameters are
${A, B, C, D} = {0.57, 0.19, 0.19, 0.05}$. For a given RMAT graph at
scale $N$, the number of vertices $n$ is $2^N$; the number of edges
$m$, after making the graph undirected by edge doubling, is
$2^N \cdot 32$. However, following the Graph500 specification~\cite{Graph500:2017},
we only use $m/2 = 2^N \cdot 16$ to calculate the edge
traversal rate. % Pay extra attention to this detail when
% comparing the traversal rates between different works.
Vertex numbers are
randomized using a deterministic hashing function after edge generation.

Our implementation outputs the hop-distances from the source vertex,
instead of the BFS tree required by Graph500. The cost of building such
a tree should be low in our implementation: only the destination vertices of nn
edges, without possible delegate parents, would need to communicate their
parent information at the end of BFS; vertices visited by dd, dn,
and nd kernels can get the parent information locally, with almost no extra cost
to the local computation.

For each reported data point, we executed 140 BFS runs with randomly
generated sources; only the ones that executed for more than 1
iteration are considered. We report the geometric mean of edge
traversal rates (in the unit of Giga Traversal Edges Per Second,
GTEPS) or elapsed times (in the unit of milliseconds, ms).

% Our BFS implementation outputs the hop-distance from the source vertex, and the
% results are verified for correctness.

We use \textit{number of nodes $\times$ number of MPI ranks per node
  $\times$ number of GPUs per MPI rank} to denote the hardware for our
experiments and for prior work. For example, $4 \times 1 \times 2$
means 4 nodes with 1 MPI rank per node, and 2 GPUs per MPI rank, 8
GPUs in total.

\subsection{Parameter Settings}
\label{sec:parameters}

\begin{figure}
  \centering
  \includegraphics[width=0.49\textwidth]{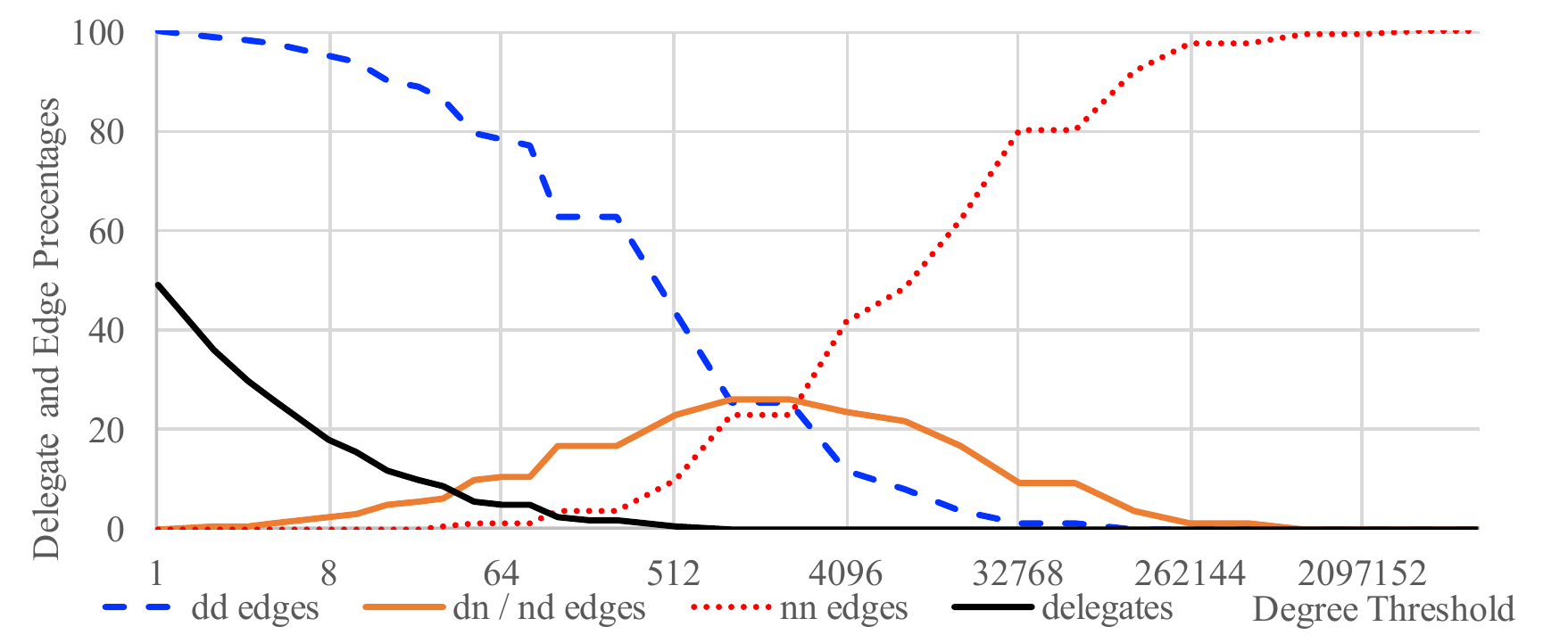}
  \centering
  \caption{Distribution of different kinds of edges and delegates,
     as a function of degree threshold, for a scale-30 RMAT graph.
     \label{fig:dsweep1}}
\end{figure}

\begin{figure}
  \centering
  \includegraphics[width=0.49\textwidth]{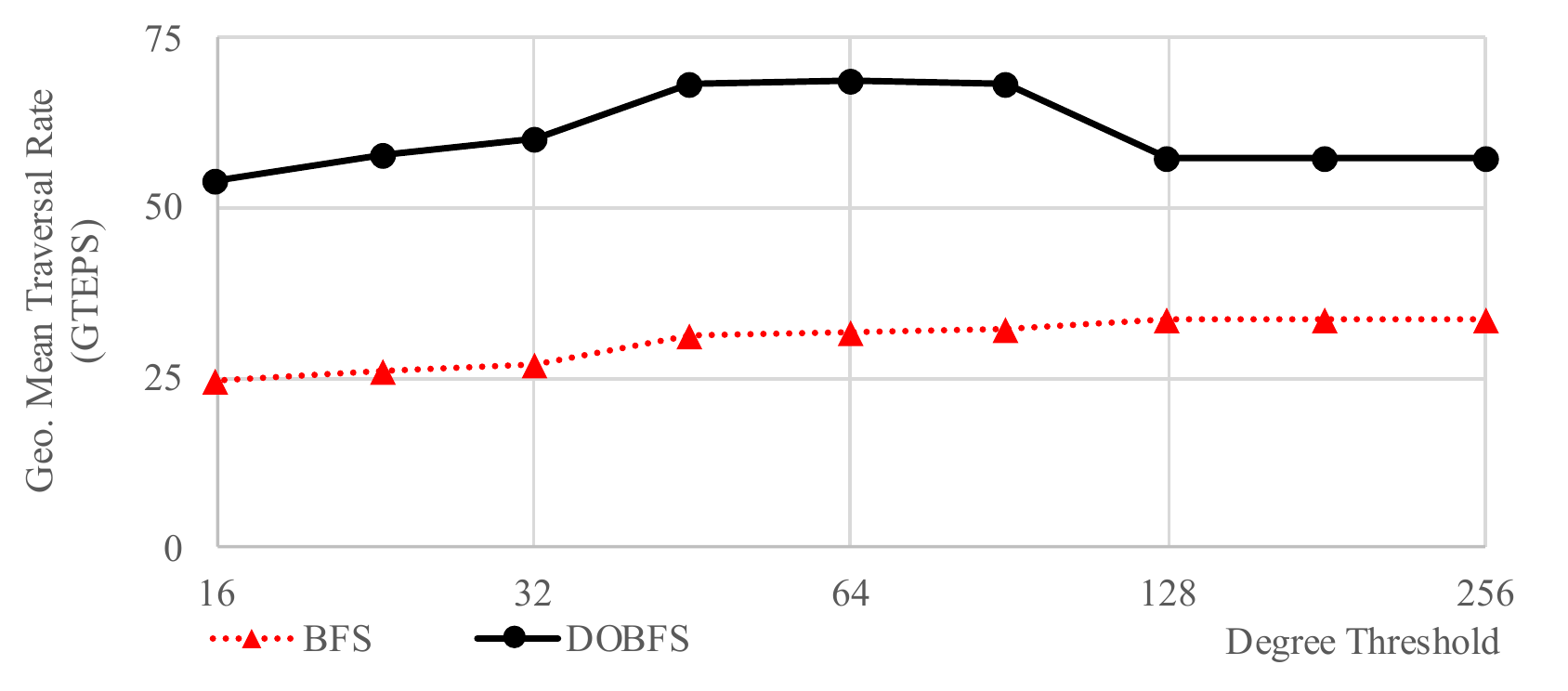}
  \centering
  \caption{Traversal rates vs.\ degree threshold,
     for a scale-30 RMAT graph with $4 \times 1 \times 4$ GPUs.
     \label{fig:dsweep2}}
\end{figure}

\begin{figure}
  \centering
  \includegraphics[width=0.49\textwidth]{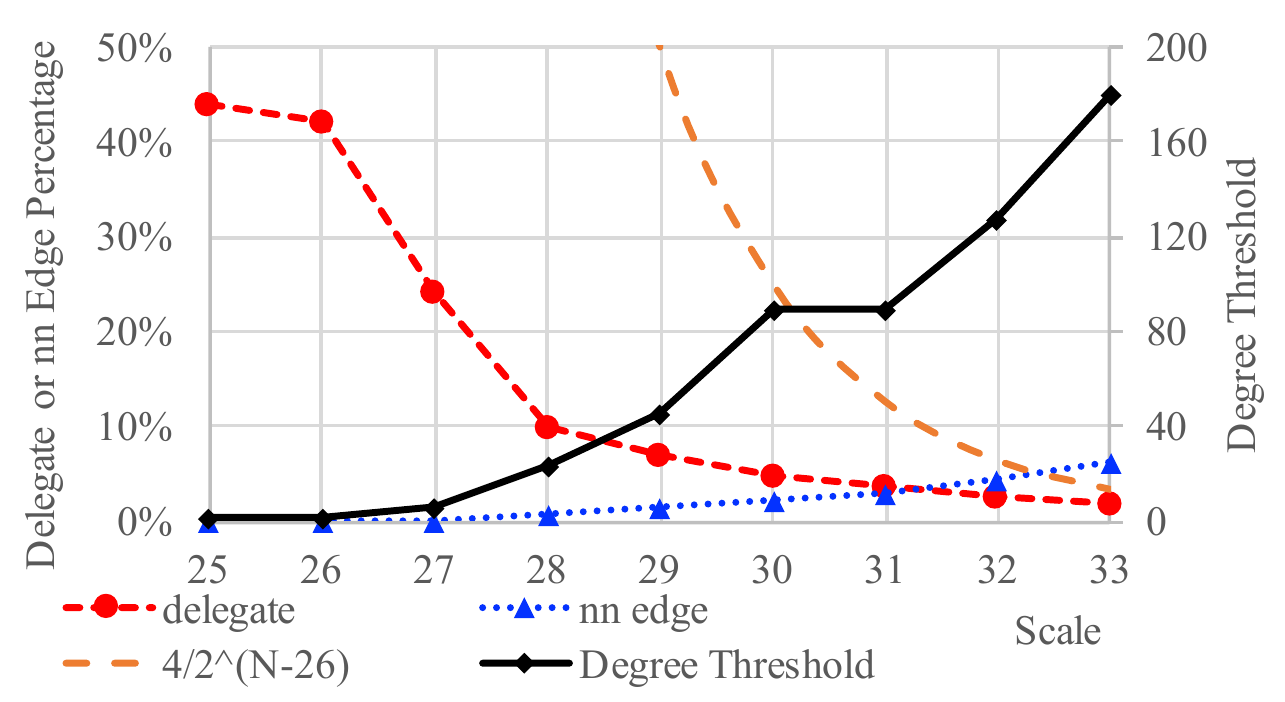}
  \centering
  \caption{Suggested degree thresholds for different RMAT scales,
    with the resulting delegate and nn edges percentages. The $4/2^{N-26}$
    line is the percentage of $4n/p$ vertices when using scale 26 RMAT
    graph per GPU\@.
    \label{fig:dsweep3}}
\end{figure}

Our implementation has several parameters and options that can be used
to tune performance. The single most important parameter is the degree
threshold \TH\@. By changing \TH, we are balancing the percentages of
delegates and nn edges. Generally we want $d$ to be on the same order
as the number of local vertices $n/p$; in our experiment, we keep $d$ under
$4n/p$. It would also be desirable to keep the nn edge percentage under
10\%. Figure~\ref{fig:dsweep1} shows how \TH{} changes the
distributions of vertices and edges on the scale-30 RMAT graph. Any
\TH{} in the range of $\lbrack 16, 512 \rbrack$ will satisfy our goal.
We sweep this range to see the resulted performance, as shown in
Figure~\ref{fig:dsweep2}. The actual range that gives the best
performance for both BFS and DOBFS is quite wide, from 45 to 90; we
use 64 in our experiments.

With a similar experiment, we suggest degree thresholds for a wide
range of graph scales (Fig.~\ref{fig:dsweep3}). The optimal \TH{}
increases at the rate of about $\sqrt{2}$ per scale. For scales
up to 33, the delegate percentage is well below the $4n/p$ line; at
scale 33, the delegate percentage is 1.75\%, and the $4n/p$ line
is at 3.23\%. The nn edge percentages increases
slightly, to 6.3\% at scale 33, which is still a small and acceptable
percentage. For larger scales that may lead to insufficient GPU memory
caused by a large number of delegates or nn edges, the following
options may be considered: 1) increase \TH{} to
decrease the number of delegates, as a range of values yield similar
performance; 2) increase $p$ to reduce the memory usage
per GPU, as there is no limitation on how many GPUs can be used,
provided that the GPU memory is sufficient. With these two options, we
believe our method could continue to scale on larger GPU clusters.

\begin{figure}
  \centering
  \includegraphics[width=0.49\textwidth]{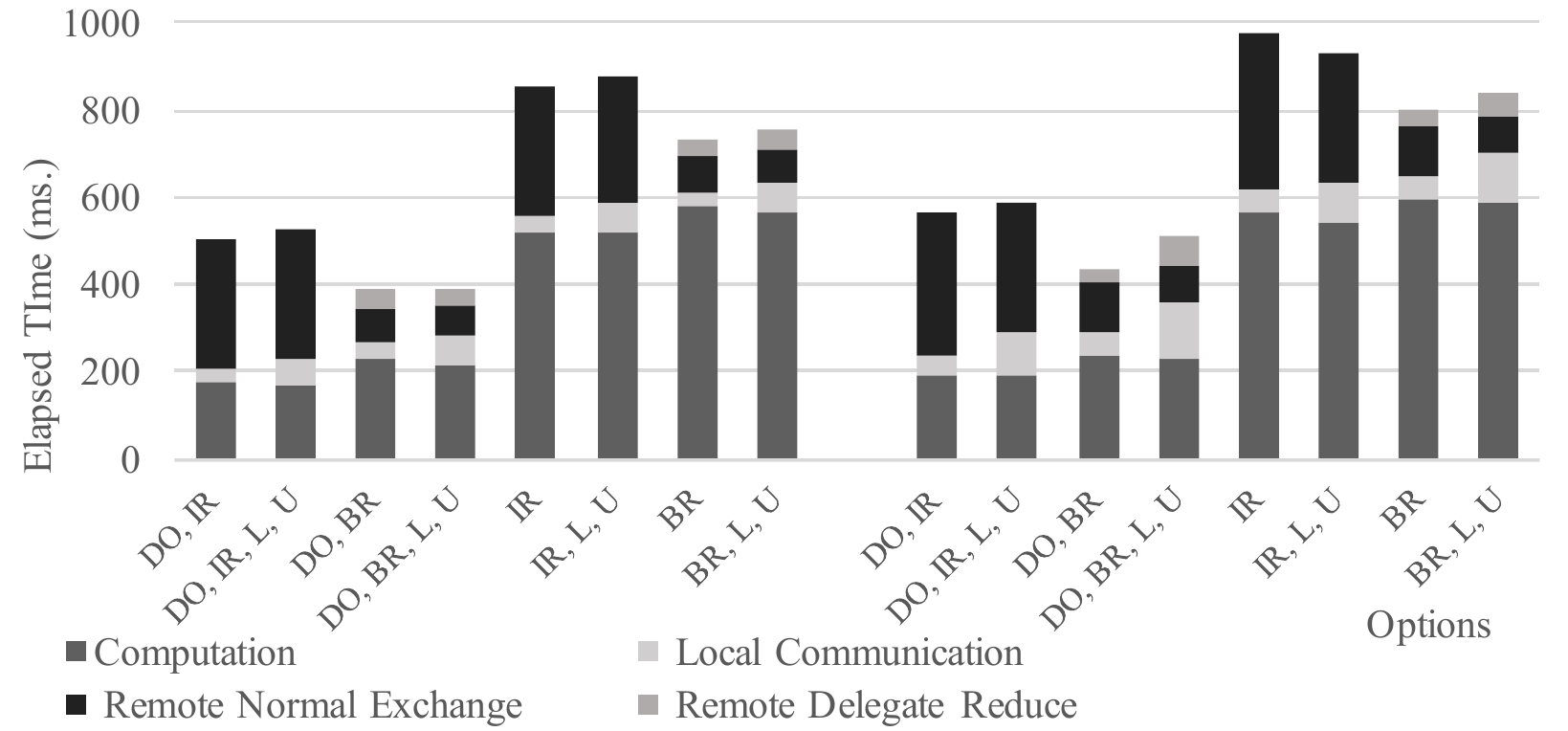}
  \centering
  \caption{Effect of different options on performance. DO stands for directional
  optimization; L for local-all2all; U for uniquify; IR for unblocking delegate
  mask reduction; and BR for blocking reduction. The graph is RMAT
  scale 32 with degree threshold 128, with a $16 \times 2 \times 2$
  hardware configuration on the
  left and $16 \times 1 \times 4$ on the right.\label{fig:options}}
\end{figure}

We can tune our implementation with several options: directional
optimization (DO), local all2all (L), uniquify (U), blocking global
mask reduction (BR) using \textit{MPI\_Allreduce} or unblocking
reduction (IR) using \textit{MPI\_Iallreduce}, and hardware
configuration (e.g., a $*\times 2 \times 2$ or $* \times 1 \times 4$
setup). Figure~\ref{fig:options} shows how different options affect
the timings of different parts of the BFS runtime. DO cuts the
computation time by a factor of three, even when the workload is
distributed on 64 GPUs. L and U add a small amount of time to local
data exchange, but do not have a significant impact on the global
communication time, mainly because the degree threshold \TH{} is so low
that we see few duplications in the normal vertex exchange. BR
significantly reduces the communication time in this example, although
the actual volumes of communication are the same. This may be a
consequence of an unoptimized implementation of
\textit{MPI\_Iallreduce}, a newly available feature on this machine.
When running on fewer than 8 nodes, the communication time of IR is
less than that of BR. We hope the same applies to a larger number of
nodes so that the advantage of workload overlapping can be fully
explored. The sum of all parts in one column is more than the elapsed
time of BFS, because different parts may overlap. For example,
visiting from the delegates can start once the delegate masks are
received without waiting for the normal vertices. For this particular
experiment, the overlaps reduce the running time by about 10\% on average
when compared to the sum of all parts.

%\begin{figure}
%  \centering
%  \includegraphics[width=0.49\textwidth]{fig/DOSweep.pdf}
%  \centering
%  \caption{DOBFS traversal rates over the change of direction
%    switching factors, for a scale-30 RMAT graph on 16 GPUs. Without DO,
%    the traversal rate is 32.0 GTEPS\@.
%    \label{fig:dosweep}}
%\end{figure}

For each of the three subgraphs that apply DO, our implementation has
two direction-switching factors that decide when to change the traversal
direction. For RMAT, once the traversal switches to the backward
direction, it does not need to change back; as a result, we only have
three factors to decide.
%and Figure~\ref{fig:dosweep} shows how they
%affect the performance.
After scanning these factors from $10^{-8}$ to $10$ for the best performance,
we found out that all three factors have a wide range of
near-optimal values; in fact, the same range
($0.5, 0.05, 1\times 10^{-7}$) for dd, dn, and nd subgraphs applies to
almost all configurations that follow the weak scaling curve and the
suggested \TH{} values. From our experience, these selections are
similar for the same type of graphs.

\subsection{Overall Results and Comparisons}

\begin{figure}
  \centering
  \includegraphics[width=0.49\textwidth]{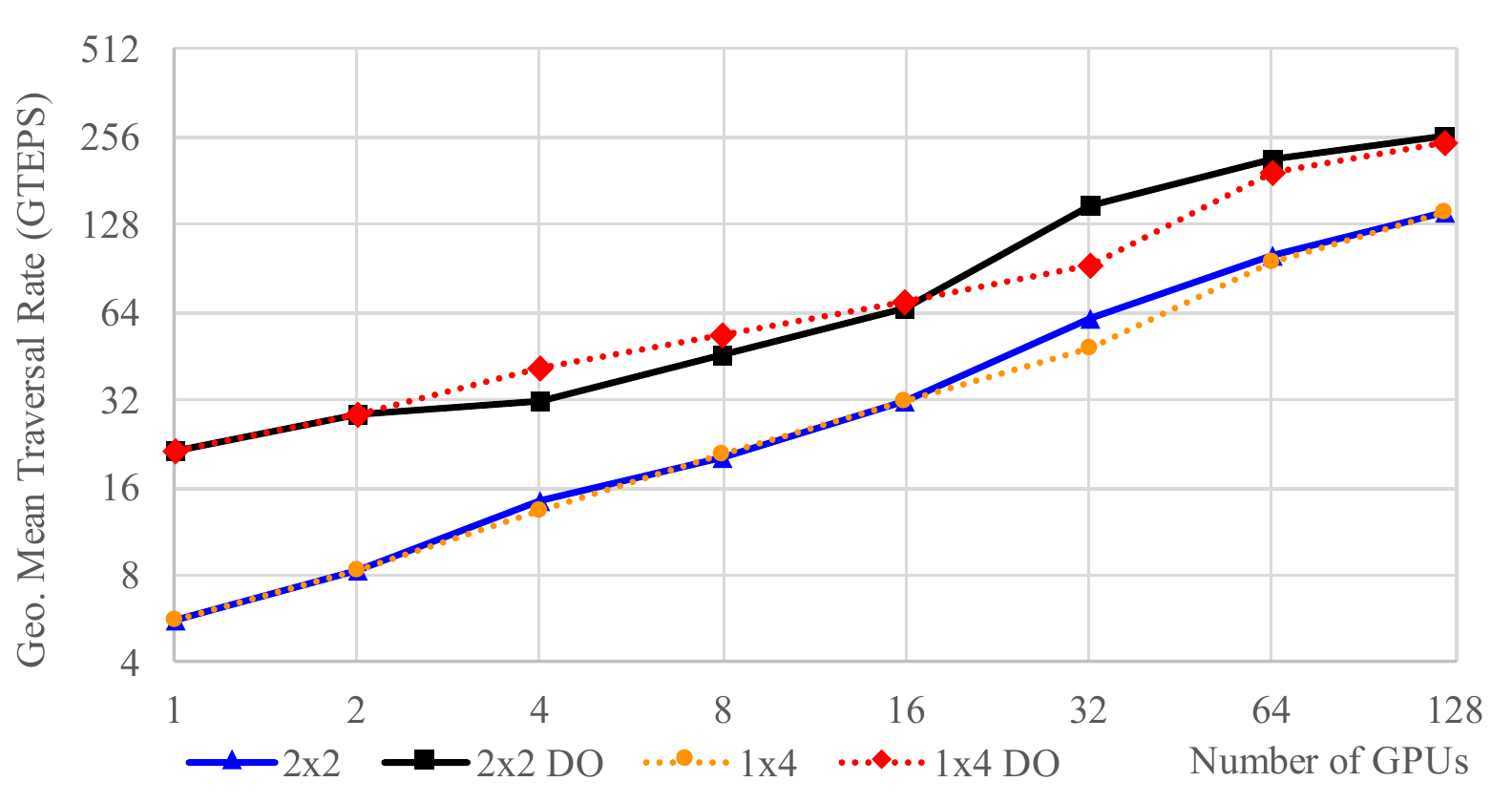}
  \centering
  \caption{Weak scaling with a scale-26 RMAT graph per GPU
     \label{fig:weakscale}}
\end{figure}

Figure~\ref{fig:weakscale} shows overall weak scaling curves, with
$\sim$scale-26 RMAT graphs on each GPU up to 124 GPUs. In this range it is
mostly linear, peaking at 259.8~GTEPS for RMAT scale 33 on 124 GPUs.
From 16 GPUs to 32 GPUs, we switch from \textit{MPI\_Iallreduce} to
\textit{MPI\_Allreduce}, as discussed earlier in this section, which
introduces performance increases higher than the average.

\begin{figure}
  \centering
  \includegraphics[width=0.49\textwidth]{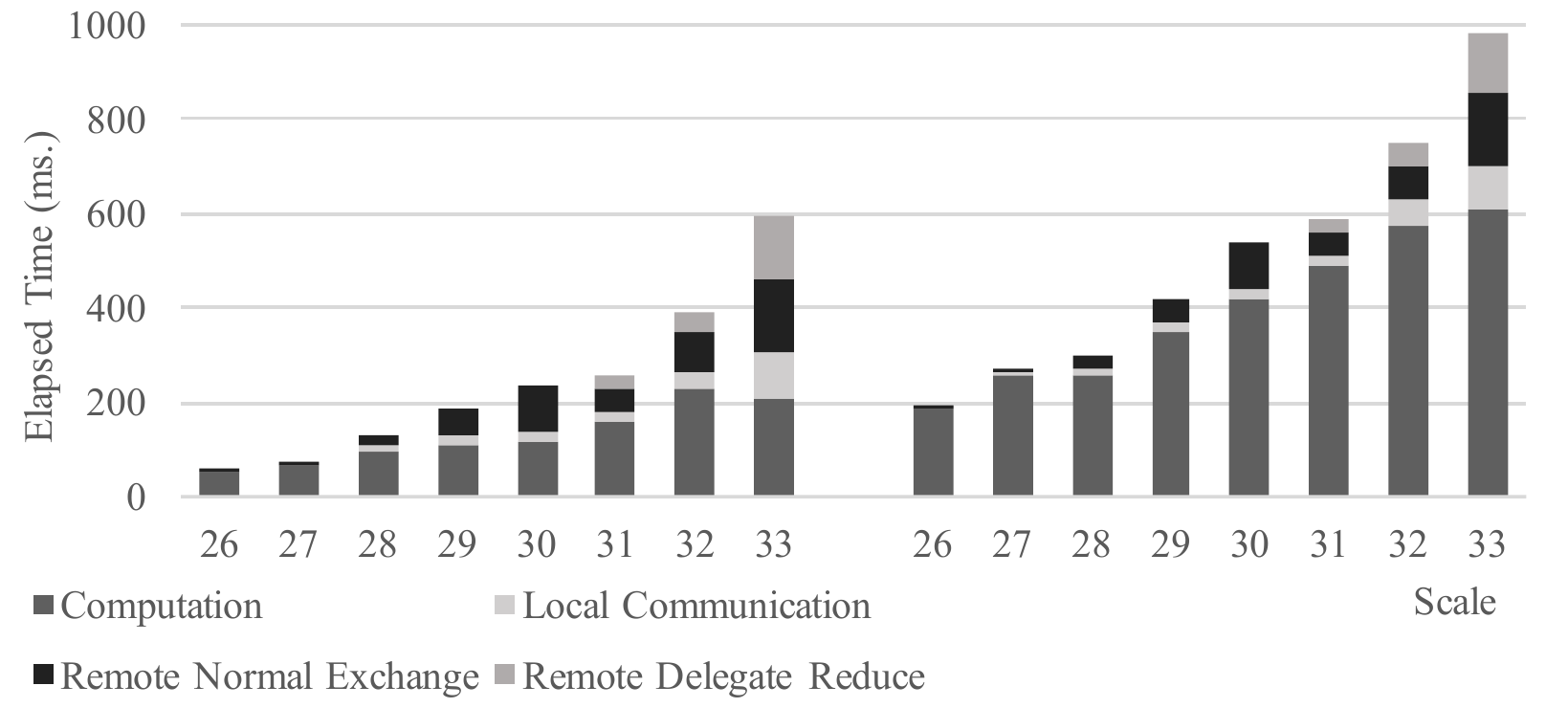}
  \centering
  \caption{Runtime breakdown for $*\times 2 \times 2$ setup along the weak
  scaling curve. DOBFS is on the left, and BFS on the right; scales 28 to 30 use
  unblocking global delegate mask reductions and merge communication time for
  masks and normal vertices; scale 31 to 33 use blocking global
  delegate mask reduction. Because of overlap, the sum of different
  parts in a column is not equal to the BFS running time.
  \label{fig:analysis}}
\end{figure}

Figure~\ref{fig:analysis} shows detailed timing for DOBFS and BFS at
different scales. Local visiting time grows slowly, only $4\times$
over 7 scales for DOBFS as the graph size and the number of GPUs
increase to $124\times$. The BFS computation time increases to
$3\times$ for the same range. This shows the computation is scaling as
expected. The communication grows slightly faster than the
computation, especially from scale 32 to 33. This may be caused by the
increases in number of delegates and nn edges, as shown previously in
Figure~\ref{fig:dsweep1} and Section~\ref{sec:parameters}, or it may
be traffic conditions in the network, as about 70\% of
the nodes in the cluster are actively transmitting large amounts of data.
Because our
implementation overlaps communication and computation, we mitigate the
effects of this increase in communication cost. % For the particular run
% of scale 33 DOBFS shown, the actual BFS elapsed time is 538.8 ms, and
% it is 10\% less than the sum of all parts.
Both computation and communication appear to successfully scale
throughout this range of RMAT sizes.

\begin{figure}
  \centering
  \includegraphics[width=0.49\textwidth]{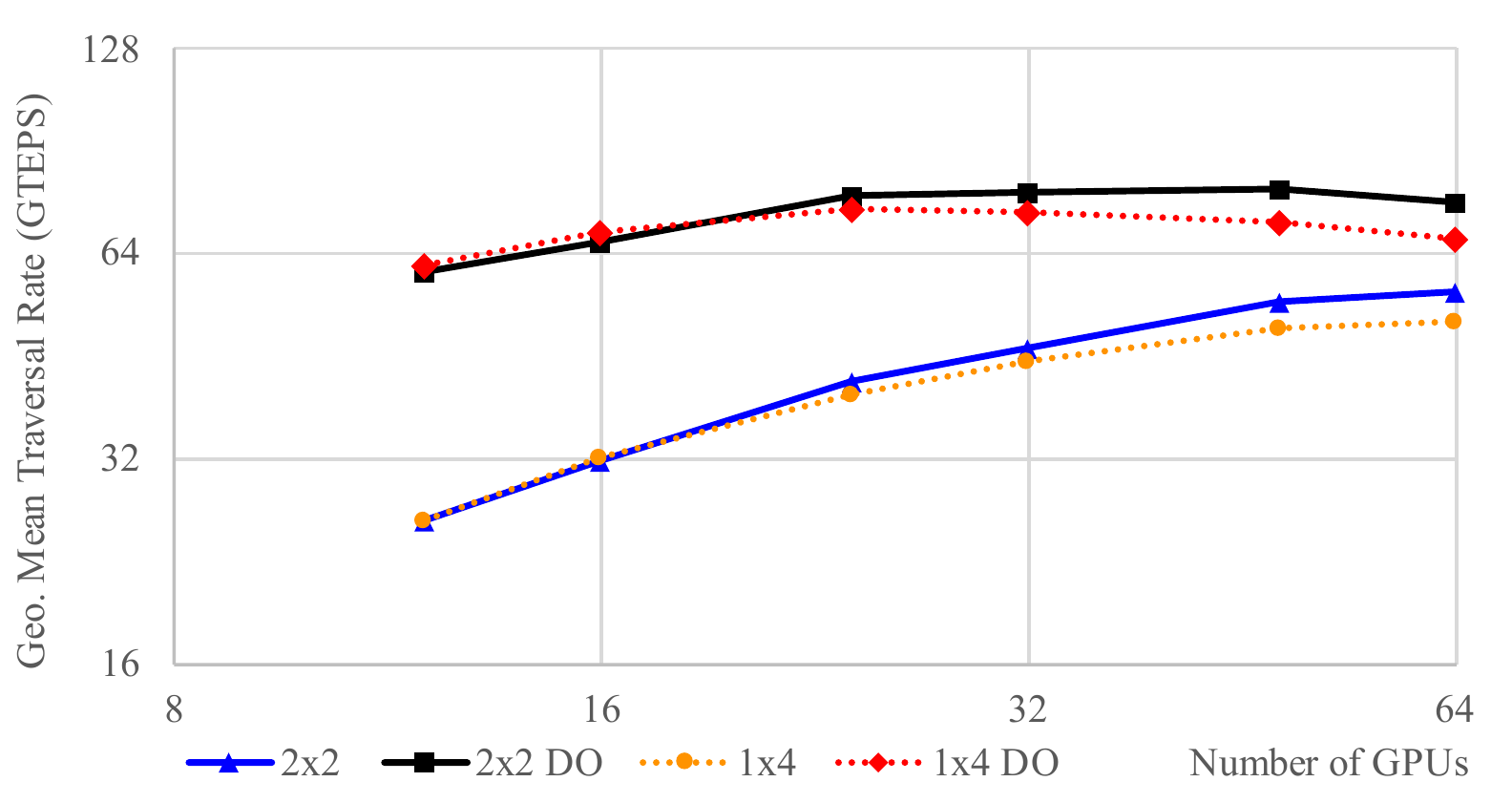}
  \centering
  \caption{Strong scaling with a scale-30 RMAT graph.
     \label{fig:strongscale}}
\end{figure}

Figure~\ref{fig:strongscale} shows strong scaling curves using the scale-30 RMAT
graph. Because of our efficient graph representation, we can fit the 34 billion
edge graph onto 12 GPUs, at about 2.9 billion edges per GPU\@. The performance
of DOBFS increases 29\% when using 24 GPUs instead of 12, and the strong scaling
curve stays almost flat after 24 GPUs. When using more GPUs, the timing
improvement in computation is about the same as the increase in communication,
caused by delegate masks reduction across more nodes and more cutting nn edges.
On more than 48 GPUs, the communication time is dominant and the GPUs are
under-utilized, thus the performance starts to drop. BFS yields better strong
scalings than DOBFS,  primarily because of its comparably larger computation
workload.

\begin{table*}[t]
  \centering
    \resizebox{\textwidth}{!}{\begin{tabular}{@{}ccccccc@{}}
      \toprule
      scale & ref. & ref.\ hw. & ref. comm. & ref.\ perf. & our hw. & our perf.\\
      \midrule

      \{24, 25, 26\} & Pan~\cite{Pan:2017:MGA:nourl}
      & 1$\times$1$\times$\{1, 2, 4\} Tesla P100 & single node & \{31.6, 42.9, 46.1\}
      & 1$\times$1$\times$\{1, 2, 4\} Tesla P100 & \{22.9, 32.5, 39.8\} \\

      33 & Bernaschi~\cite{Bernaschi:2015:EGD}
      & 4096$\times$1$\times$1 Tesla K20X & Dragonfly 100Gbps & 828.39
      & 31$\times$2$\times$2 Tesla P100 & 259.8 \\

      29 & Krajecki~\cite{Krajecki:2016:BTM}
      & 64$\times$1$\times$1 Tesla K20Xm & FatTree 10Gbps & 13.7
      & 2$\times$1$\times$4 Tesla P100 & 53.13 \\

      33 & Yasui~\cite{Yasui:2017:FSE}
      & 128$\times$10$\times$1/10 Xeon E5-4650 v2& shared memory & 174.7
      & 31$\times$2$\times$2 Tesla P100 & 259.8 \\

      33 & Buluc~\cite{Buluc:2017:DMB}
      & 1204$\times$1$\times$1 Xeon E5-2695 v2& Dragonfly 64Gbps & $\sim$240
      & 31$\times$2$\times$2 Tesla P100 & 259.8 \\

      \bottomrule
     \end{tabular}}
     \caption {Comparison with previous works.
     \label{tab:compare}}
\end{table*}

We compare our results with previous efforts in
Table~\ref{tab:compare}. When compared against single-node multi-GPU
Gunrock~\cite{Pan:2017:MGA:nourl}, this work is a little slower when using
the same graphs, which may be the effect of more optimizations in
Gunrock's traversal kernels. As we add more GPUs in this work, we see
the gap in performance is narrowing, which indicates better
scalability; and the memory size improvements we made in this paper
allows us to process larger graphs on one node, up to scale 28 on 4
GPUs, than any other GPU-based previous work.

Compared to Bernaschi et al.~\cite{Bernaschi:2015:EGD}, our work
achieves about 31\% of their performance with only 3\% the number of
GPUs. Although the GPUs they used are not as new as ours, the
10$\times$ per-GPU performance shows our efficient computation and
communication. Compared to Krajecki, Loiseau, Alin, and Jaillet~\cite{Krajecki:2016:BTM},
we achieve 4$\times$ the performance using only one eighth the number
of GPUs.

The flagship shared-memory CPU implementation by Yasui and Katsuki~\cite{Yasui:2017:FSE} uses
a similar number of processors; we obtained 1.49$\times$ the
performance of their work, which we believe is partially because of
the performance advantages of the GPU\@. We also demonstrate slightly
better performance than Bulu\c{c} et al.~\cite{Buluc:2017:DMB} despite
their 8.4$\times$ more processors.

\subsection{General Graphs and Applications}

\begin{figure}
  \centering
  \includegraphics[width=0.49\textwidth]{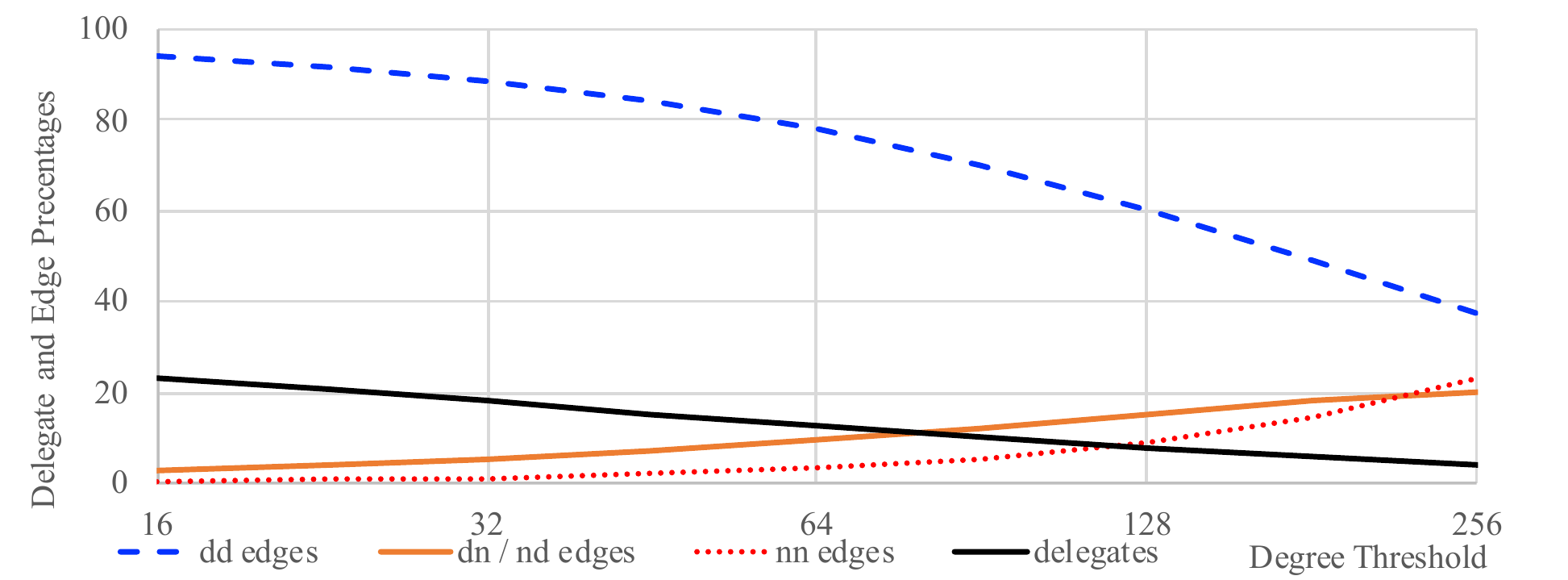}
  \centering
  \caption{Distribution of different kinds of edges and delegates, as a function
     of degree threshold, for the friendster graph.
     \label{fig:dsweep4}}
\end{figure}

\begin{figure}
  \centering
  \includegraphics[width=0.49\textwidth]{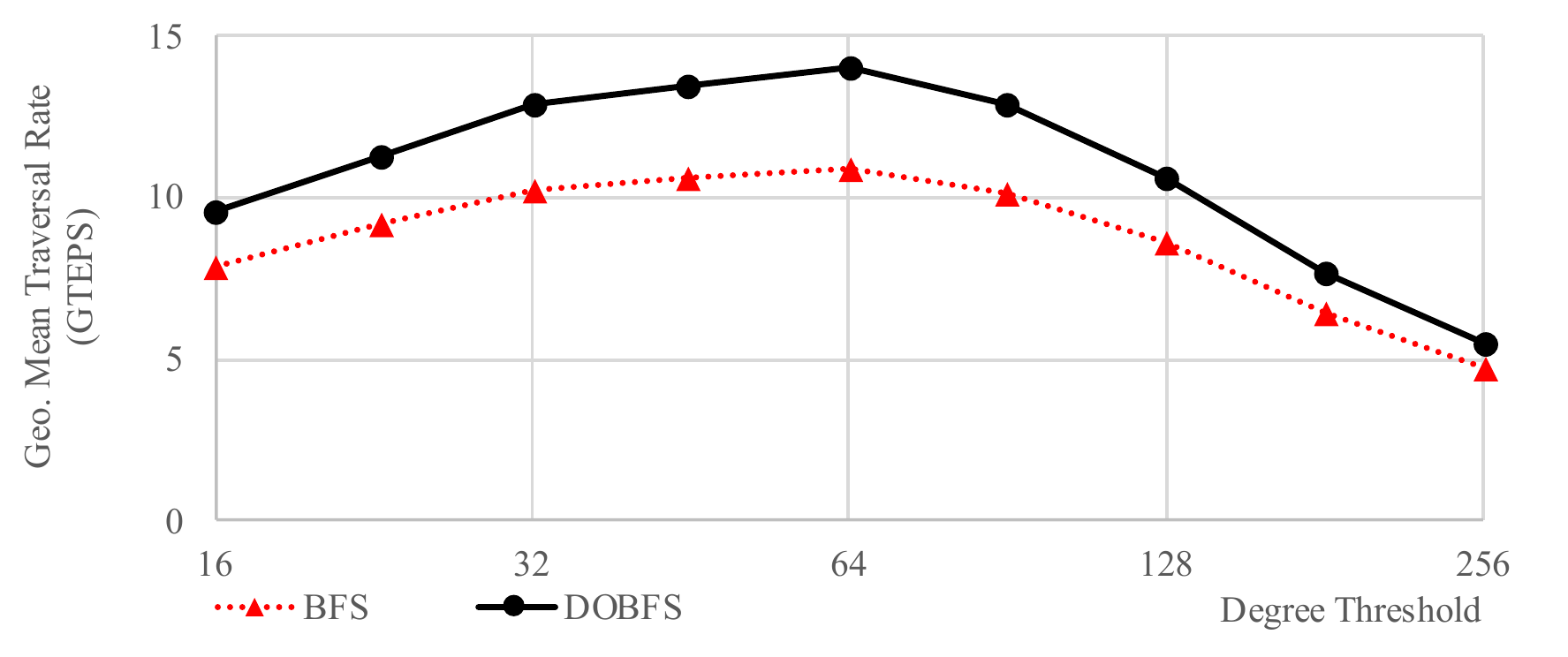}
  \centering
  \caption{Traversal rates vs. degree threshold, for the friendster graph
     with $ 1 \times 2 \times 2$ GPUs.
     \label{fig:dsweep5}}
\end{figure}

We are also interested in graphs more general than RMAT, and use the Friendster graph~\cite{Friendster}
to test our implementation. We prepare this graph by randomizing the vertex numbers and make the graph
symmetric by edge doubling. The resulting graph has 134 million vertices, about
half of which are isolated ones, and 5.17 billion edges. Figure~\ref{fig:dsweep4}
shows how the distribution of different kinds of edges and vertices change with
regard to various degree thresholds \TH, and Fig.~\ref{fig:dsweep5} shows the
resulting performance. Similar to RMAT, the friendster social network has a wide
range of suitable \TH{} values, in $\lbrack 16, 128 \rbrack$. There is also a
wide range $\lbrack 32, 91 \rbrack$ of \TH{} values that gives close to the best
performance.

We also analyze the Web Data Common (WDC) 2012 hyperlink graph~\cite{WDC2012},
but reduce the hyper edges into single ones. Vertex-randomization and  edge-doubling
give a graph with 4.29 billion vertices, 402 million of which are zero-degree ones,
and 224 billion edges. Using 160 GPUs in a $40 \times 2 \times 2$ setup, our
implementation achieves 84.2 GTEPS and 79.7 GTEPS, for BFS and DOBFS respectively,
both using degree threshold 256. The BFS searches have about 330 iterations at
average, and experience long-tail behavior. The averaged per-iteration time
of 8~$\mu$S is not much more than the per-iteration overhead of a few $\mu$S\@.
In this situation, the additional workload for direction decisions in DOBFS is
more than the workload saving in traversal. As a result, the DOBFS performance
is slightly less than BFS\@.

For algorithms more general than BFS, in most cases, the local computation is
at least in the order of $O(m)$, much larger than that of DOBFS\@. They also
introduce larger communication volumes: BFS only needs 1 bit for the visited
status of delegates, and 32 bits for newly visited normal vertices of cutting
nn edges across GPUs. Other graph algorithms require more bits of state for
delegates---for example, ranking scores for PageRank---and associative values
for normal vertices in additional to the vertex numbers themselves. For large
scale-free graphs, the increases in computation and communication are roughly
in the same order, and our computation and communication models should still
be scalable. For graph processing that yields insufficient local workloads
over many iterations, either caused by the graph topologies or the algorithms,
we argue that they may not be suitable for Bulk Synchronous Parallel (BSP)
frameworks on systems with fat nodes: the GPUs will be underutilized, and the
per-iteration overhead may well make such implementations unscalable. Asynchronous
graph frameworks, such as HavoqGT~\cite{HavoqGT} and Groute~\cite{Ben-Nun:2017:GAA},
may be more suitable for such workloads.

\section{Conclusions}
\label{sec:conclusion}
Base on the idea of separating vertices by out-degrees, we implemented a
scalable BFS, consisting of an efficient graph representation, scalable and
fast local computation kernels, and a scalable communication model. With 124
P100 GPUs on the CORAL EA system, we achieved 259.8~GTEPS on the scale 33 RMAT
graph.
The close-to-linear weak scaling indicates that our work successfully
targets modern GPU clusters, which feature fewer nodes and more local
computing power than previous systems.

We believe our work
%\footnote{This work was performed, in part, under the auspices of the U.S.
%Department of Energy by Lawrence Livermore National Laboratory
%under Contract DE-AC52-07NA27344.  Experiments were performed at the
%Livermore Computing facility.}
 provides a better alternative to conventional 2D
partitioning methods for scaling DOBFS, and is better aligned with the
latest trend of supercomputers and large systems. Further exploration
using even more GPUs in the range of thousands, when they are available,
could bring us more insight into and solutions for the
scalability problem.

Future work includes investigating graph applications beyond
BFS\@. These applications need more local computation than just neighborhood
queries, more communication than just 1-bit visited status, and
more attributes on vertices and edges than a single label.
While most techniques and optimizations for BFS should still be applicable,
we hope to see further work in the components of graph representation, local
computation, and remote communication, under more complex application scenarios.

\section*{Acknowledgments}
\label{sec:ack}
 This work was performed, in part, under the auspices of the U.S.
 Department of Energy by Lawrence Livermore National Laboratory
 under Contract DE-AC52-07NA27344 (LLNL-CONF-742180).  Experiments were performed at the
 Livermore Computing facility. We appreciate the support of NSF grants
 CCF-1629657 and OAC-1740333, the Defense Advanced Research Projects
 Agency (DARPA), and an Adobe Data Science Research Award.

 We intend to open source our code from this work. Please contact rpearce@llnl.gov for the latest updates.
\bibliographystyle{IEEEtran}
\bibliography{bib/all,temp}

\begin{thebibliography}{10}
\providecommand{\url}[1]{#1}
\csname url@samestyle\endcsname
\providecommand{\newblock}{\relax}
\providecommand{\bibinfo}[2]{#2}
\providecommand{\BIBentrySTDinterwordspacing}{\spaceskip=0pt\relax}
\providecommand{\BIBentryALTinterwordstretchfactor}{4}
\providecommand{\BIBentryALTinterwordspacing}{\spaceskip=\fontdimen2\font plus
\BIBentryALTinterwordstretchfactor\fontdimen3\font minus
  \fontdimen4\font\relax}
\providecommand{\BIBforeignlanguage}[2]{{%
\expandafter\ifx\csname l@#1\endcsname\relax
\typeout{** WARNING: IEEEtran.bst: No hyphenation pattern has been}%
\typeout{** loaded for the language `#1'. Using the pattern for}%
\typeout{** the default language instead.}%
\else
\language=\csname l@#1\endcsname
\fi
#2}}
\providecommand{\BIBdecl}{\relax}
\BIBdecl

\bibitem{Graph500:2017}
``The {J}une 2017 {Graph500} list,'' \url{https://graph500.org/?page_id=254},
  Jun. 2017.

\bibitem{Buluc:2008:OTR}
A.~Bulu{\c{c}} and J.~R. Gilbert, ``On the representation and multiplication of
  hypersparse matrices,'' in \emph{Proceedings of the 22nd IEEE International
  Symposium on Parallel and Distributed Processing Symposium}, ser. IPDPS 2008,
  Apr. 2008, pp. 1--11.

\bibitem{Agarwal:2010:SGE}
V.~Agarwal, F.~Petrini, D.~Pasetto, and D.~A. Bader, ``Scalable graph
  exploration on multicore processors,'' in \emph{Proceedings of the
  International Conference for High Performance Computing, Networking, Storage
  and Analysis}, ser. SC '10, Nov. 2010, pp. 46:1--46:11.

\bibitem{Beamer:2012:DBS}
S.~Beamer, K.~Asanovi\'{c}, and D.~Patterson, ``Direction-optimizing
  breadth-first search,'' in \emph{Proceedings of the International Conference
  on High Performance Computing, Networking, Storage and Analysis}, ser. SC
  '12, Nov. 2012, pp. 12:1--12:10.

\bibitem{Pan:2017:MGA:nourl}
Y.~Pan, Y.~Wang, Y.~Wu, C.~Yang, and J.~D. Owens, ``Multi-{GPU} graph
  analytics,'' in \emph{Proceedings of the 31st IEEE International Parallel and
  Distributed Processing Symposium}, ser. IPDPS 2017, May\slash Jun. 2017, pp.
  479--490.

\bibitem{CORAL}
``Coral info,'' Oct. 2017, \url{https://asc.llnl.gov/coral-info}.

\bibitem{Ray}
``{Ray - Livermore Computing},'' Oct. 2017,
  \url{https://hpc.llnl.gov/hardware/platforms/Ray}.

\bibitem{Pearce:2014:FPT}
R.~Pearce, M.~Gokhale, and N.~M. Amato, ``Faster parallel traversal of scale
  free graphs at extreme scale with vertex delegates,'' in \emph{Proceedings of
  the International Conference for High Performance Computing, Networking,
  Storage and Analysis}, ser. SC '14, Nov. 2014, pp. 549--559.

\bibitem{Yasui:2017:FSE}
Y.~Yasui and K.~Fujisawa, ``Fast, scalable, and energy-efficient parallel
  breadth-first search,'' in \emph{The Role and Importance of Mathematics in
  Innovation: Proceedings of the Forum ``Math-for-Industry'' 2015}.\hskip 1em
  plus 0.5em minus 0.4em\relax Springer Singapore, 2017, pp. 61--75.

\bibitem{Vastenhouw:2005:TDD}
B.~Vastenhouw and R.~H. Bisseling, ``A two-dimensional data distribution method
  for parallel sparse matrix-vector multiplication,'' \emph{{SIAM} Review},
  vol.~47, no.~1, pp. 67--95, 2005.

\bibitem{Liu:2015:EBG}
H.~Liu and H.~H. Huang, ``Enterprise: Breadth-first graph traversal on
  {GPU}s,'' in \emph{Proceedings of the International Conference for High
  Performance Computing, Networking, Storage and Analysis}, ser. SC '15, Nov.
  2015, pp. 68:1--68:12.

\bibitem{Ben-Nun:2017:GAA}
T.~Ben-Nun, M.~Sutton, S.~Pai, and K.~Pingali, ``Groute: An asynchronous
  multi-{GPU} programming model for irregular computations,'' in
  \emph{Proceedings of the 22nd ACM SIGPLAN Symposium on Principles and
  Practice of Parallel Programming}, ser. PPoPP '17, Feb. 2017, pp. 235--248.

\bibitem{Maass:2017:MPT}
S.~Maass, C.~Min, S.~Kashyap, W.~Kang, M.~Kumar, and T.~Kim, ``Mosaic:
  Processing a trillion-edge graph on a single machine,'' in \emph{Proceedings
  of the 12th European Conference on Computer Systems}, ser. EuroSys 2017, Apr.
  2017, pp. 527--543.

\bibitem{Ueno:2016:ESB}
K.~Ueno, T.~Suzumura, N.~Maruyama, K.~Fujisawa, and S.~Matsuoka, ``Extreme
  scale breadth-first search on supercomputers,'' in \emph{Proceedings of the
  2016 IEEE International Conference on Big Data}, ser. BigData 2016, Dec.
  2016, pp. 1040--1047.

\bibitem{Lin:2017:SGT}
H.~Lin, X.~Tang, B.~Yu, Y.~Zhuo, W.~Chen, J.~Zhai, W.~Yin, and W.~Zheng,
  ``Scalable graph traversal on {S}unway {T}aihu{L}ight with ten million
  cores,'' in \emph{Proceedings of the 31st IEEE International Parallel and
  Distributed Processing Symposium}, ser. IPDPS 2017, May\slash Jun. 2017, pp.
  635--645.

\bibitem{Buluc:2017:DMB}
A.~Bulu{\c{c}}, S.~Beamer, K.~Madduri, K.~Asanovic, and D.~A. Patterson,
  ``Distributed-memory breadth-first search on massive graphs,'' \emph{CoRR},
  vol. abs/1705.04590, 2017.

\bibitem{Ueno:2013:PDB}
K.~Ueno and T.~Suzumura, ``Parallel distributed breadth first search on
  {GPU},'' in \emph{Proceedings of the 20th International Conference on High
  Performance Computing}, ser. HiPC 2013, Dec. 2013, pp. 314--323.

\bibitem{Bernaschi:2015:EGD}
M.~Bernaschi, G.~Carbone, E.~Mastrostefano, M.~Bisson, and M.~Fatica,
  ``Enhanced {GPU}-based distributed breadth first search,'' in
  \emph{Proceedings of the 12th ACM International Conference on Computing
  Frontiers}, ser. CF '15, May 2015, pp. 10:1--10:8.

\bibitem{Fu:2014:PBF}
Z.~Fu, H.~K. Dasari, B.~Bebee, M.~Berzins, and B.~Thompson, ``Parallel breadth
  first search on {GPU} clusters,'' in \emph{Proceedings of the 2014 IEEE
  International Conference on Big Data}, ser. BigData 2014, Oct. 2014, pp.
  110--118.

\bibitem{Krajecki:2016:BTM}
M.~Krajecki, J.~Loiseau, F.~Alin, and C.~Jaillet, ``{BFS} traversal on
  multi-{GPU} cluster,'' in \emph{Proceedings of the 2016 IEEE International
  Conference on Computational Science and Engineering}, ser. CSE 2016, Jul.
  2016, pp. 594--599.

\bibitem{Young:2016:OCF}
J.~Young, J.~Romera, M.~Hauck, and H.~Fr{\"{o}}ning, ``Optimizing communication
  for a 2{D}-partitioned scalable {BFS},'' in \emph{Proceedings of the 2016
  IEEE High Performance Extreme Computing Conference}, ser. HPEC 2016, Sep.
  2016, pp. 1--7.

\bibitem{Green:2016:CSD}
O.~Green and D.~A. Bader, ``{cuSTINGER}: Supporting dynamic graph algorithms
  for {GPU}s,'' in \emph{Proceedings of 2016 IEEE High Performance Extreme
  Computing Conference}, ser. HPEC 2016, Sep. 2016, pp. 1--6.

\bibitem{Sakharnykh:2016:BGM}
N.~Sakharnykh, ``Beyond {GPU} memory limits with unified memory on {P}ascal,''
  Dec. 2016,
  \url{https://devblogs.nvidia.com/parallelforall/beyond-gpu-memory-limits-unified-memory-pascal/}.

\bibitem{Davidson:2014:WPG:nourl}
A.~Davidson, S.~Baxter, M.~Garland, and J.~D. Owens, ``Work-efficient parallel
  {GPU} methods for single source shortest paths,'' in \emph{Proceedings of the
  28th IEEE International Parallel and Distributed Processing Symposium}, ser.
  IPDPS 2014, May 2014, pp. 349--359.

\bibitem{Merrill:2012:SGG}
D.~Merrill, M.~Garland, and A.~Grimshaw, ``Scalable {GPU} graph traversal,'' in
  \emph{Proceedings of the 17th ACM SIGPLAN Symposium on Principles and
  Practice of Parallel Programming}, ser. PPoPP '12, Feb. 2012, pp. 117--128.

\bibitem{NVIDIA:2014:NNH}
{NVIDIA Corporation}, ``{NVIDIA} {NVLink} high-speed interconnect: Application
  performance,'' NVIDIA Corporation, Tech. Rep., Nov. 2014,
  \url{http://info.nvidianews.com/rs/nvidia/images/NVIDIA%20NVLink%20High-Speed%20Interconnect%20Application%20Performance%20Brief.pdf}.

\bibitem{Friendster}
``Friendster social network dataset: Friends,'' Jul. 2011,
  \url{https://archive.org/details/friendster-dataset-201107}.

\bibitem{WDC2012}
``Web data commons - hyperlink graphs,'' 2012,
  \url{http://webdatacommons.org/hyperlinkgraph/}.

\bibitem{HavoqGT}
R.~Pearce, ``{HavoqGT},'' 2015, \url{https://github.com/LLNL/havoqgt}.

\end{thebibliography}

\end{document}